\RequirePackage{lineno}
\documentclass[aps,showpacs,preprintnumbers,amsmath]{revtex4}
\usepackage{graphicx}
\usepackage{dcolumn}
\usepackage{epstopdf}
\usepackage{color}
\usepackage{comment}
\usepackage{subfigure}

\usepackage{bm}
\begin{document}
	\title{Quantifying the Underlying Event: \\ Investigating Angular Dependence of Multiplicity Classes and Transverse-momentum Spectra in High-energy pp Collisions at LHC Energies}
	\author{Aditya Nath Mishra{$^{1}$},  Gergely G\'abor Barnaf\"oldi{$^{1}$} and Guy Pai{\'c}{$^2$}}
	\medskip
	
	\affiliation{$^{1}$Wigner Research Centre for Physics, 29-33 Konkoly-Thege Mikl\'os str., 1121 Budapest, Hungary \\
	$^{2}$Instituto de Ciencias Nucleares, Universidad Nacional Aut\'onoma de M\'exico, \\ Apartado Postal 70-543, M\'exico City 04510, M\'exico}
	
	\bigskip
	\date{\today}

\begin{abstract}
We present a study of the transverse momentum spectra and their evolution in function of the position of the azimuthal of the particles associated to the leading particle. Additionally, the behavior of the spherocity distribution in the same azimuthal bins is reported. The studies were made using  proton-proton collisions at $\sqrt{s}$ = 13 TeV using PYTHIA8 Monte Carlo event generator. The Multiplicity and midrapidity transverse momentum spectra of charged hadrons have been analyzed in the non-extensive statistical framework. The results on the findings corresponding to the Underlying Event are reported.
\end{abstract}

\maketitle

\section{Introduction}
\label{intro}
Numerous experimental measurements over a wide range of energies and collision systems have been devoted to studying the properties of the QCD matter formed in ultra-relativistic nuclear collisions. A number of measurements have provided strong indications that the matter formed in the heavy-ion collisions behaves as a strongly interacting fluid that exhibits collectivity~\cite{Arsene:2005,BBBack:2005,Adams:2005,Adcox:2005,Chatrchyan:2012,Abelev:2013,Aad:2013}.
Initially, small collision systems, like proton-proton and proton-nucleus, have been used to study initial and final state effects in “cold” nuclear matter, treating them as a baseline for the interpretation of heavy-ion results. However,  recently, collective, fluid-like features strikingly similar to those observed in heavy-ion collisions, have been also observed in small collisions  systems  especially in high-multiplicity events~\cite{Cole:2007}. This suggests, that in compressed, dense matter, the correlation length is getting shorter, thus the number of degrees of freedom are increased.

Strongly correlated statistical systems can be investigated by correlation methods. However, while angular correlations with triggers refer on the momentum space, the modern analytical tools and the high statistics allows us to study more in detail the distributions in phase space and the associated entropy production.
The use of event structure tools like the spherocity can also be applied.
Finite and strongly correlated systems have been investigated recently in the non-extensive statistical framework. Applying the Tsallis-entropy~\cite{Biro:2020kve} and its connection to the high-energy nuclear collisions, opened new windows by using the so called Tsallis\,--\,Pareto distribution~\cite{Biro:2020kve}. The Tsallis parameters of the transverse momentum ($p_T$) distributions of the produced hadron states provide a better description of the full spectra than other functional forms. The distribution, which can be derived from the Tsallis-entropy formula~\cite{Biro:2020kve}, fits well both at low- and high $p_T$, better than either thermal or perturbative QCD models. Indeed, the spectra in the intermediate transverse momentum range (2 GeV/$c$ $\lesssim p_T \lesssim 6$ GeV/$c$), is well described, providing a unified description between the bulk, hydro-like, and the perturbative ranges. Another aspect of the non-extensive statistical framework is that the spectrum parameters: the temperature, $T$ and non-extensivity, $q$, carry physical information to interpret the system created in the high-energy collision. In a recent work, 
the parameter-evolution of Tsallis\,--\,Pareto transverse momentum distributions has been identified with high predictive power at various event multiplicities. This presents a new view of the equation of state of the high temperature and extremely dense nuclear matter, which can be investigated in geometrical and event shape aspects as well as by angular-correlation triggers.

In this work, the aim is to qualitatively and quantitatively classify high-energy events in comparison to the well known event shape observable. The method here is to generate hadron spectra in ultra-relativistic proton-proton collisions, extract the Tsallis-parameters in given multiplicity classes, which provide an entropy-based event classification. This led us to understand the strongly correlated, non-perturbative regime and to determine the absolute and quantitative properties of the Underlying Event (UE).

The structure of this work is as follows: we present the event generator settings and the simulated data. We explain our method and we define global quantities and the two cases for our correlation analysis. Global and event shape variables are given, by fitting the hadron spectra in the angular bins, where Tsallis\,--\,Pareto parameters, multiplicity values and transverse spherocities are determined. Introducing different angular-correlation slices as cases, allow us to compare model parameters with each other and with various theoretical approaches. 
We also provide the Tsallis-thermometer for angular regions. In a final step, the correlation of these parameters with multiplicity and event-shape variables are carried out and cross checked for consistency.
\section{The Simulated Data}
\label{pythia}

PYTHIA~\cite{pythia6,pythia8.1,pythia8.2} is a general-purpose Monte Carlo event generator, widely used for the generation of events in high-energy collisions of leptons, protons and nuclei. It has undergone decades of development and tuning to collider and other data. The event generation in PYTHIA consists of several steps starting typically from a hard scattering process, followed by initial and final state parton showering, multiparton interactions, and the final hadronization process.

The results reported in this paper are obtained from 1 billion non-diffractive events for pp collisions at $\ensuremath{\sqrt{s}}$ = 13 TeV simulated using PYTHIA version 8.240 with the default Monash 2013 tune~\cite{pythiaMonash}.

The events and particles are selected for this analysis according to the following criteria.
The events having at least three primary charged particle with transverse momentum $p_T > $ 0.15 GeV/$c$ within the pseudorapidity $|\eta| <$ 0.8 are analyzed, which selection is required by the later-defined event-shape variable calculations.
The primary charged particles are defined as all final state particles including decay products except coming form weak decays of strange particles, this definition is similar to the one used by the ALICE experiment~\cite{alice:2017}, thus might be considered for experimental investigations in the future.  
\section{Analysis Method and Definitions}
\label{Def}
In this section we present the definitions, tools, and observables used further in the paper.
\subsection{Leading Particles}
\label{LeadPar}
The particle with maximum transverse momentum in a particular event is considered the leading particle. The azimuthtal angle reflects the angle of all associated charged particles with respect to the leading particle. 
\begin{figure}[hbt]
	\centering        
	\vspace*{-0.2cm}
	\includegraphics[scale=0.5]{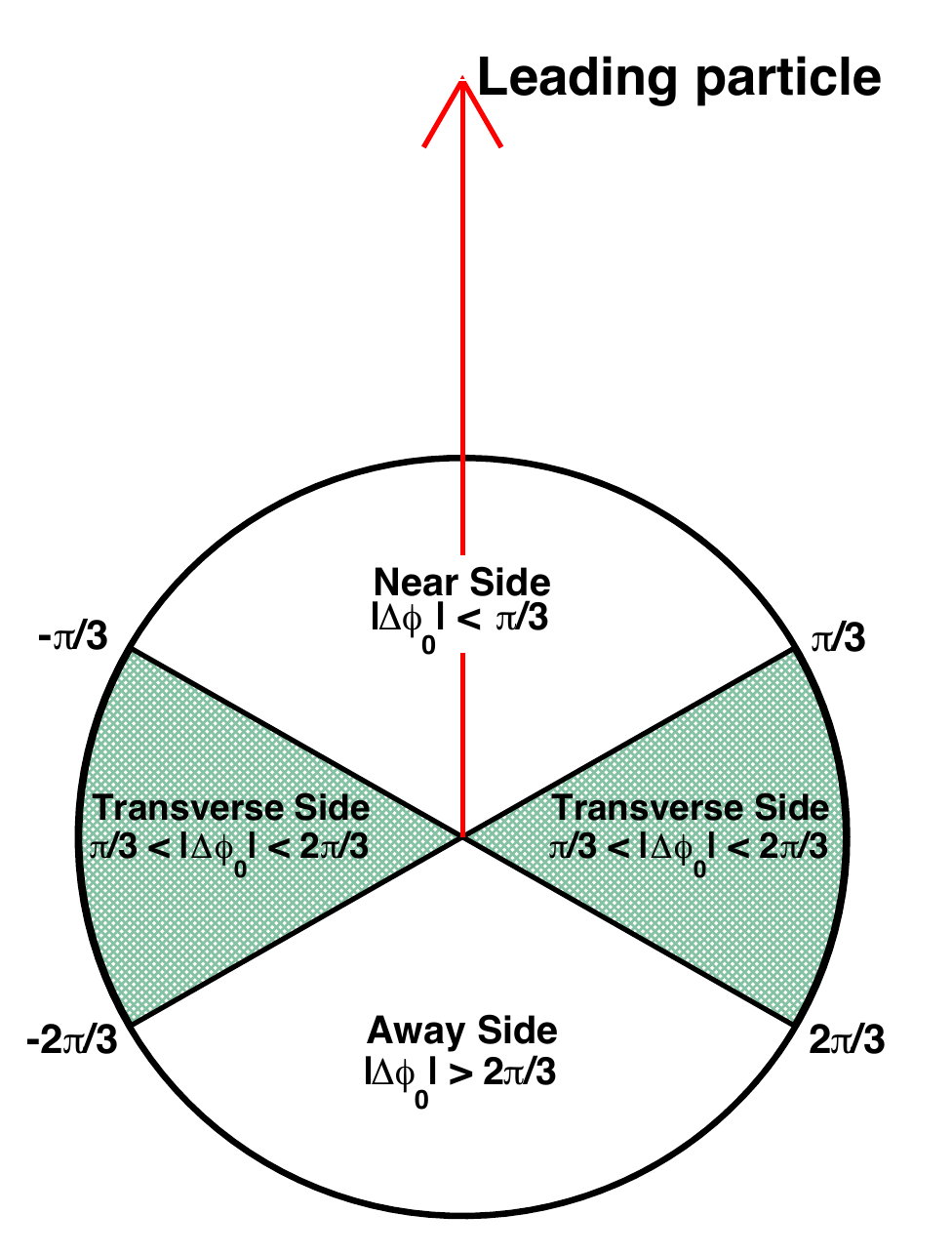}
	\vspace*{-0.5cm}
	\caption{The usual definition of regions in the azimuthal angle with respect to the leading (highest-$p_T$) charged particle, where leading charged particle is highlighted by a red arrow.}
	\label{fig1:UE}
\end{figure}
\subsection{Underlying Event}
\label{underlyingevent}
The Underlying Event (UE) consists of particles that accompany a hard scattering such as the beam-beam remnants and particles produced through Multiple-Parton Interactions (MPI). Initial-State Radiation (ISR) and Final-State Radiation (FSR) contribute to the UE. As illustrated in Fig~~\ref{fig1:UE}, the azimuthal angular difference between leading charged particle and associated charged particles of the event, $\Delta\phi_{0} = (\phi^{leading} - \phi^{assoc.})$, is used to define three distinct topological regions following the CDF Collaboration~\cite{CDFPaper1,CDFPaper2}:
\begin{itemize}
	\item Near Side (NS): $|\Delta\phi_{0}| < \pi/3 $, ($|\Delta\phi_{0}| < 60^\circ $) 
	\item Transverse Side (TS): $\pi/3 < |\Delta\phi_{0}| < 2\pi/3 $, ($ 60^\circ <|\Delta\phi_{0}| < 120^\circ $)
	\item Away Side (AS): $|\Delta\phi_{0}| >  2\pi/3 $, ($|\Delta\phi_{0}| < 120^\circ $)
\end{itemize}

The NS and the AS regions are dominated by string fragments originating from the hardest partonic process of the event. On the other hand, the TS is dominated largely by the contributions of the UE.

\begin{figure}[th!]
	\centering        
	\vspace*{-0.2cm}
	\includegraphics[scale=0.4]{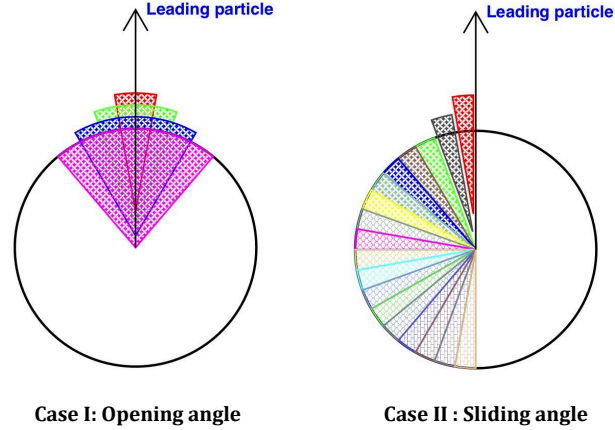}
	\vspace*{-0.5cm}
	\caption{Definitions of two Cases for $\Delta\phi$ bin selections. Case I (left): Open $\Delta \phi$ angle by 20$^\circ$, named ``opening angle". In this Case the binning starts from --10 - 10$^\circ$ and the last bin covers full azimuthal space i.e. Minimum Bias (MB). Case II (right): make slices of the $\Delta \phi$ of size 20$^\circ$, named ``sliding angle". In Case II the first bin 0 - 20$^\circ$ is treated in both in with and without leading particle scenarios.}
	\label{fig1:TwoCases}
\end{figure}

To understand the particle production mechanism, by exploring the contributions of UE activity, in the  azimuthal space  we consider two Cases, shown in Fig~\ref{fig1:TwoCases}, to divide azimuthal space, $\Delta \phi$, in 18 different sections, where $\Delta \phi$ is the angle between the leading charged particle and associated  charged particles of the event.
\begin{description}
	\item[Case I]: We open $\Delta \phi$ angle in steps of 20$^\circ$, named ``opening angle". The binning starts from --10 - 10$^\circ$ and the last bin covers full azimuthal space i.e. --180 - 180$^\circ$. It is easy to recognize, that the last bin is the Minimum Bias (MB) and, therefore the ratio of the largest $\Delta \phi$ bin to the MB found exactly one. Case I is useful to investigate the evolution of the thermodynamical observables of the system.
	
	\item[Case II]: We make slices of the $\Delta \phi$ of size 20$^\circ$, named ``sliding angle". In this case, the results for the first bin 0 - 20$^\circ$ are reported in two ways: including and excluding the leading particle in the result. Case II is a tool for exploring the geometrical structure of the Underlying Event.
	
\end{description}	

Note, Case I and Case II can be cross-checked within the 1\textsuperscript{st} bin with and without including the leading particle. This first bin is important, since this is the perturbative region, therefore well-described by theoretical models. Moreover one has to take into account, that angle opening for Case I and Case II is treated differently, which will present as different structures in the $\Delta \phi$ plots. Case I definition includes both sides of the leading particles, while Case II has mirror symmetry for $\Delta \phi = \pi$.

\subsection{Charged Hadron Multiplicity}
\label{sec:IntroMult}
The multiplicity of produced charged particles is one of the basic characteristics of high-energy hadron collisions and has been the subject of longstanding experimental and theoretical studies, which have shaped the understanding of the particle production mechanism and entropy balance. We note, the number of charged particles is calculated by the selection defined in Section~\ref{pythia}. Since we select the events having at least three primary charged particles in one of the $\Delta \phi$ bins in case second, we loose some of the events. Therefore, minimum bias reported in this paper is different than earlier studies using all events~\cite{Mishra:PRC2019,Pranjal:2021}.
\subsection{Transverse Momentum Spectra}
\label{sec:IntroSpectra}
The transverse momentum spectra of charged particles produced in high energy proton-proton collisions at the Large Hadron Collider (LHC) energies offer unique information about soft and hard interactions. The $p_T$-spectra consists of a low $p_T$-region where soft processes dominate particle production, whereas the high $p_T$-region is mostly dominated by hard processes. In this paper, we study $p_T$-spectra of charged particles using PYTHIA8 Monte-Carlo generator in pp collisions at $\sqrt{s}$ = 13 TeV to characterise the events by analysing UE contributions in the azimuthal space. 

\subsection{Event shape variable spherocity}
The mid-rapidity charged hadron transverse spherocity ($S_{0}$), hereinafter referred to as spherocity, has been used as a valuable tool to characterise the events through the geometrical distribution of the  $p_T$'s of the charged hadrons, which is by definition collinear and infrared safe~\cite{Salam:2009jx}. By restricting it to the transverse plane, transverse spherocity avoids the bias from the boost along the
beam axis~\cite{Banfi:2010xy}. Transverse spherocity is defined for a unit transverse vector $\hat{n}$ which minimizes the ratio~\cite{Cuautle:2014yda, Cuautle:2015kra}:
\begin{figure}[thbp]
	\centering        
	\vspace*{-0.2cm}
	\includegraphics[scale=0.5]{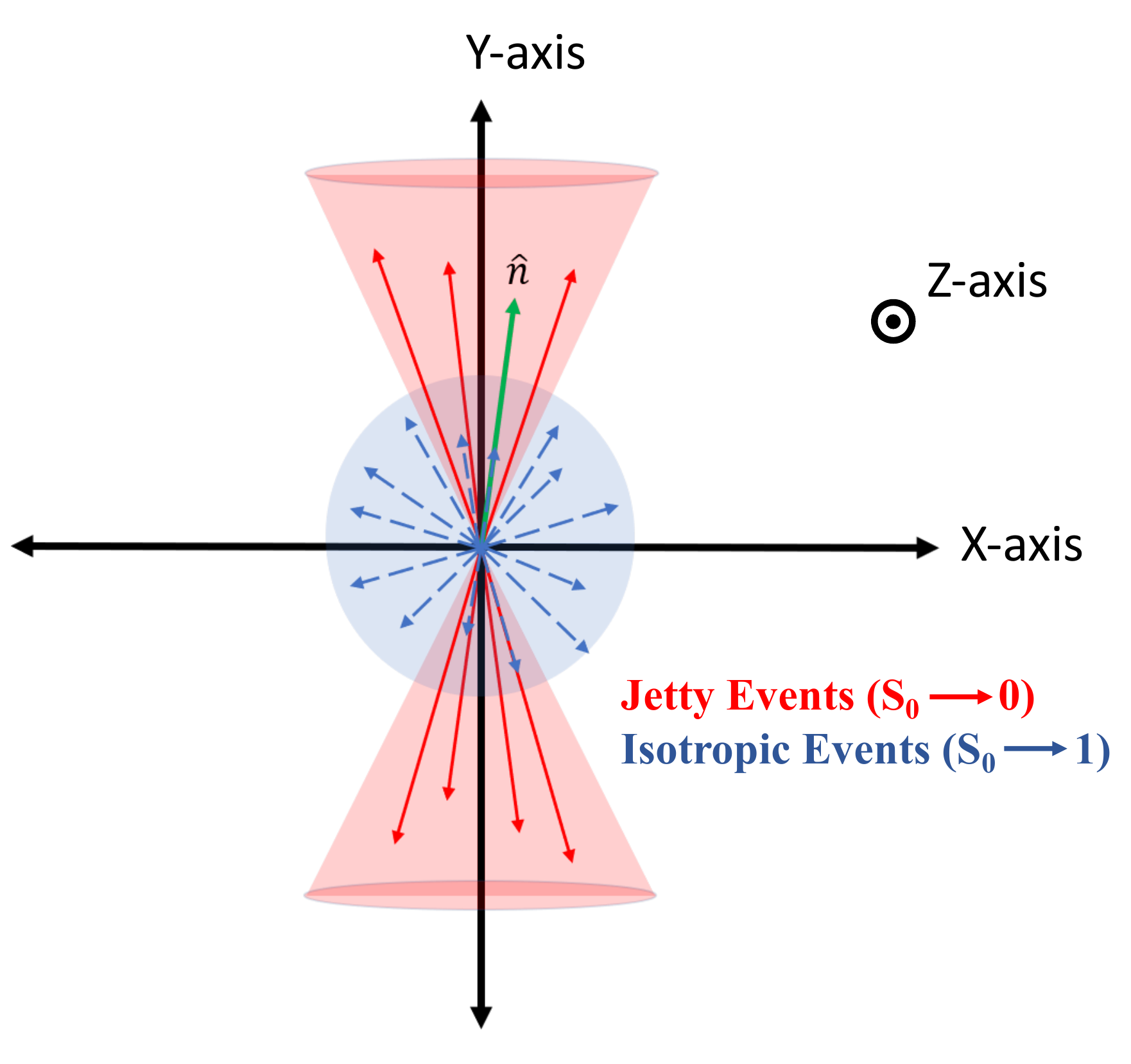}
	\vspace*{-0.5cm}
	\caption{Schematic picture of event shape variable (spherocity) showing jetty and isotropic events in the transverse plane.}
	\label{fig:S0}
\end{figure}
\begin{eqnarray}
	S_{0} = \frac{\pi^{2}}{4} \bigg(\frac{\Sigma_{i}~|\vec p_{T_{i}}\times\hat{n}|}{\Sigma_{i}~p_{T_{i}}}\bigg)^{2}.
	\label{eq1}
\end{eqnarray}
The extreme limits of $S_{0}$ are related to specific configurations of events in the transverse plane. The value of $S_{0}$ ranges from 0 to 1, which is ensured by multiplying the normalisation constant $\pi{^2}/4$ in Eq.~(\ref{eq1}). The lower limit of spherocity ($S_{0}\rightarrow0$) corresponds to event topologies where all transverse momentum vectors are (anti)parallel or the sum of the $p_T$ is dominated by a single track, called pencil like or jetty events. The upper limit of spherocity ($S_{0}\rightarrow1$) corresponds to event topologies where transverse momentum vectors are ``isotropically'' distributed, called isotropic events. The jetty events are usually the hard events while the isotropic ones are the result of soft processes. The schematic picture of the event shape variable (spherocity) showing jetty and isotropic events in the transverse plane is shown in Fig.~\ref{fig:S0}. Note, for obtaining spherocity values one needs at least three primary charged particles in the event, which was set in Section~\ref{pythia} accordingly. 

\section{Global \& Event-shape Properties}
\label{sec:eventshape}

To explore the structure and properties of events, the geometrical and momentum quantities need to be handled in parallel. Our aim is to connect the geometrical structure of multiplicity distributions and the transverse momentum spectra. These two aspects can be interconnected by the spherocity.

Although this task is well defined for a jet-like event structure, where both the momentum and the spatial dimensions are limited, this method is not unequivocal for the isotropic Underlying Event. We investigate here the angular dependence of these global observables in parallel with the event shape properties to explore the geometrical connection between them in both Case I and Case II were defined in Section~\ref{Def}.

\subsection{Multiplicity Variations with $\Delta\phi$}
\label{sec:mult}
\begin{figure*}[ht]
	\centering
	\subfigure[~Case I]{\includegraphics[scale=0.4]{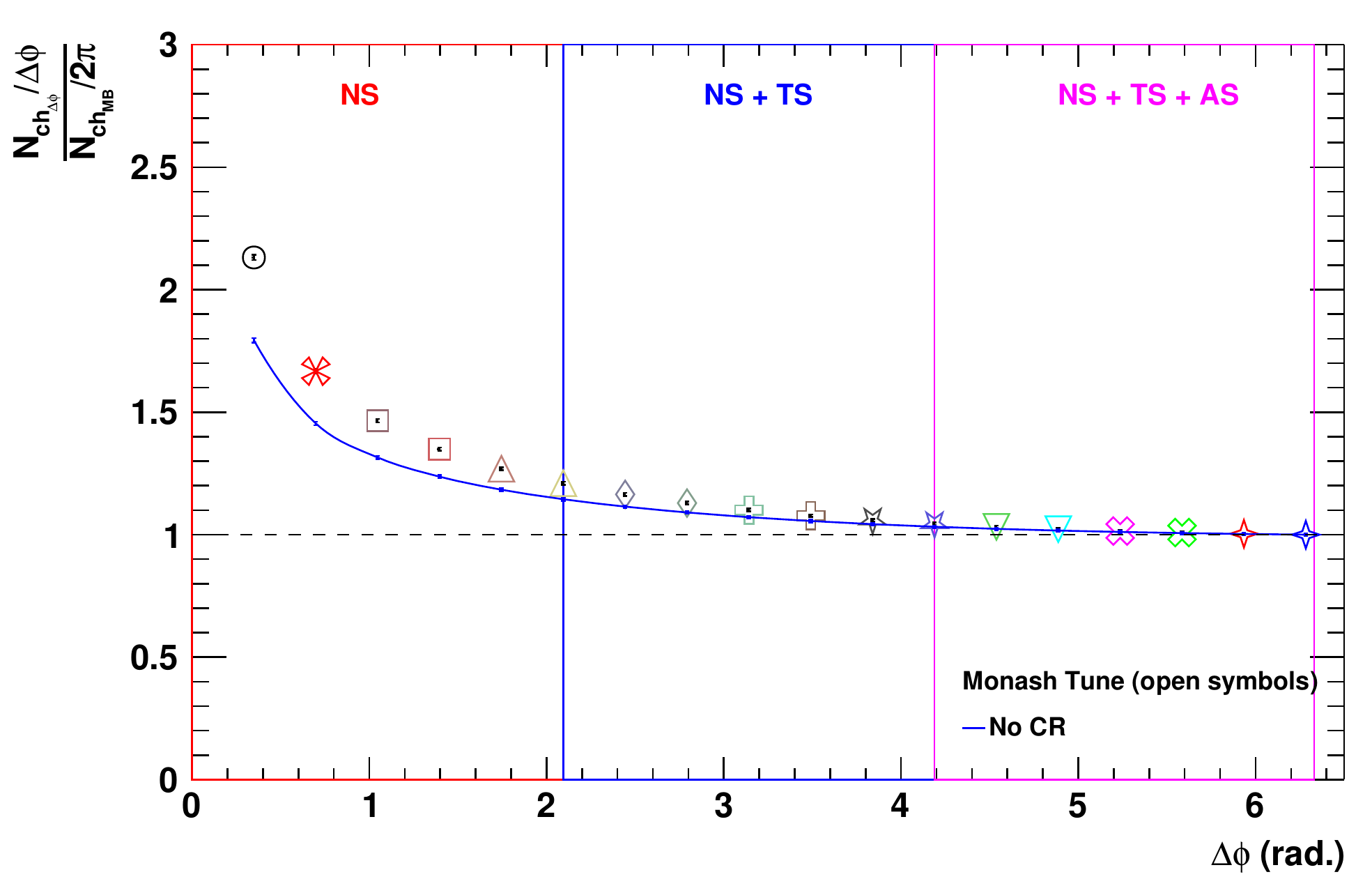}}\quad
	\label{fig3:MultOA}
	\subfigure[~Case II]{\includegraphics[scale=0.4]{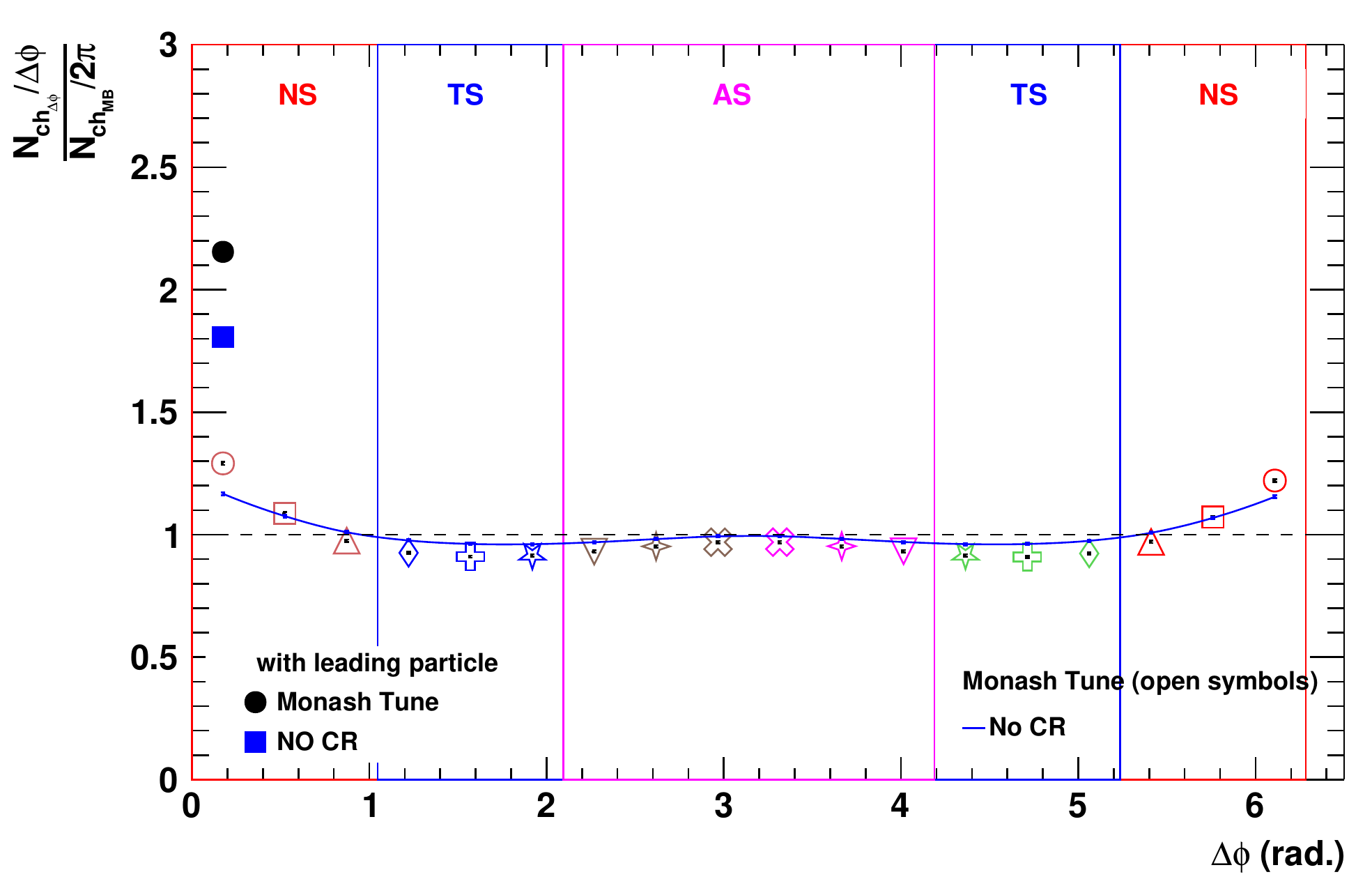}}
	\label{fig3:MultSA}
	\caption{Multiplicity in $\Delta\phi$ bin is scaled with the multiplicity from MB in the same $\Delta\phi$ bin are shown for Case I (left) and Case II (right) for both Monash tune (open symbols) and CR off (blue solid line). First bin of the Case II is plotted by full symbols for both Monash tune (black point) and CR off (blue rectangle).}
	\label{fig3:Mult}
\end{figure*}
To understand the multiplicity variation with $\Delta\phi$, we  plot charged particle multiplicity scaled it with minimum bias multiplicity ($\frac{\rm N_{\rm ch (\Delta\phi)}/\Delta\phi}{\rm N_{\rm ch (MB)}/2\pi}$) in each $\Delta\phi$ bin for both cases. Figure~\ref{fig3:Mult}(a) shows that relative multiplicity decreases with opening angle, $\Delta\phi$ and get saturated to 1 at high $\Delta\phi$ in Case I.  Here, as widening the opening angle, the hadronic degrees of freedom are increasing, which suppress the effect around the leading particle and weaken the dominance of the pencil-like (jetty) structure at the Near Side. In Fig.~\ref{fig3:Mult}(b), Case II, relative multiplicity attain value more than one in the first Near Side bin. The relative multiplicity shows a decrease from Near Side to Transverse Side and then a small rise (close to one) from Transverse Side to Away Side. In Case II the differential properties present rapidly-changing effects than in Case I as we will see later in Section~\ref{sec:ueid}.
\begin{figure*}[ht]
	\centering
	\subfigure[~Case I]{\includegraphics[scale=0.4]{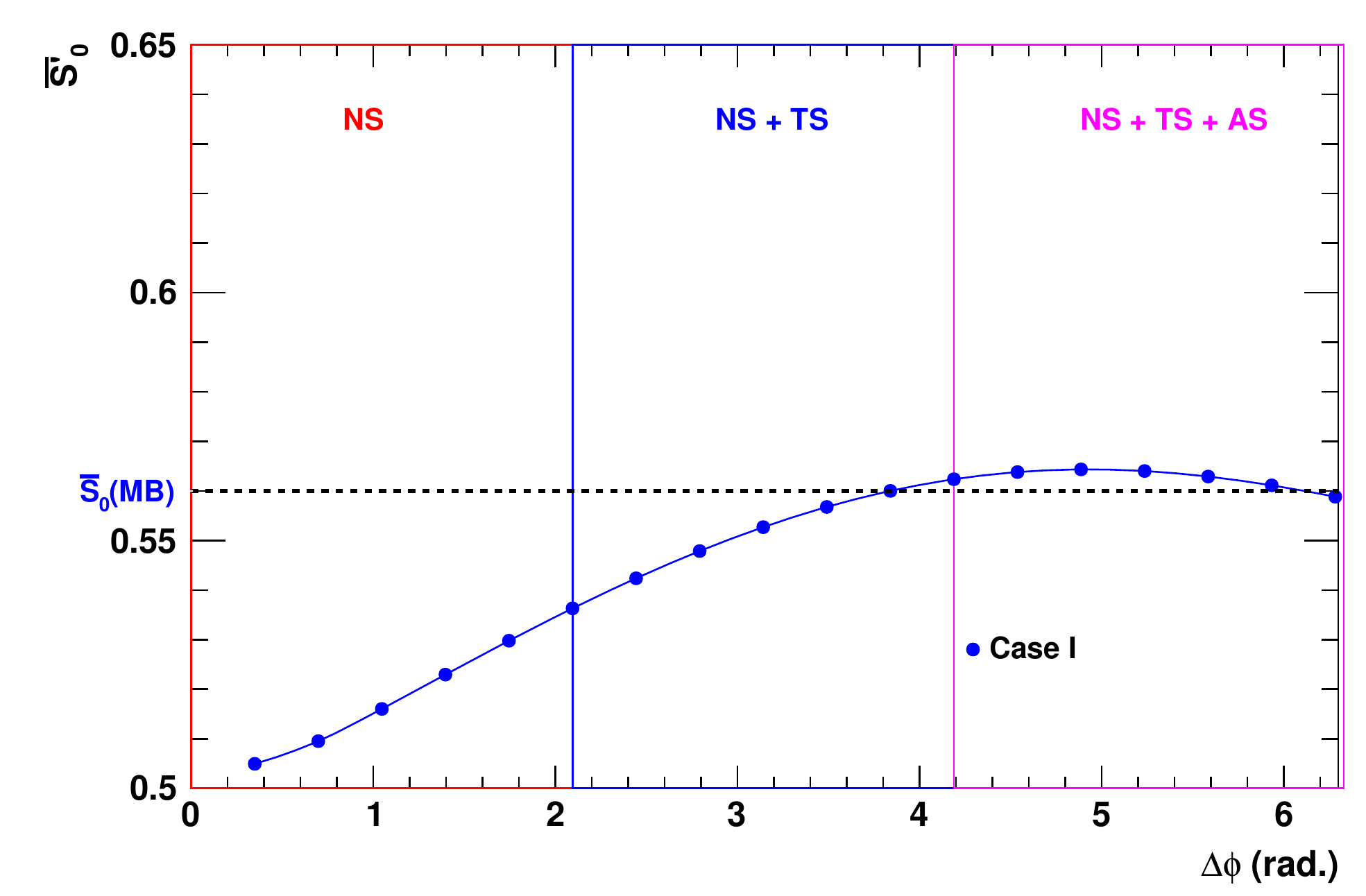}}\quad
	\label{fig:S0OA}
	\subfigure[~Case II]{\includegraphics[scale=0.4]{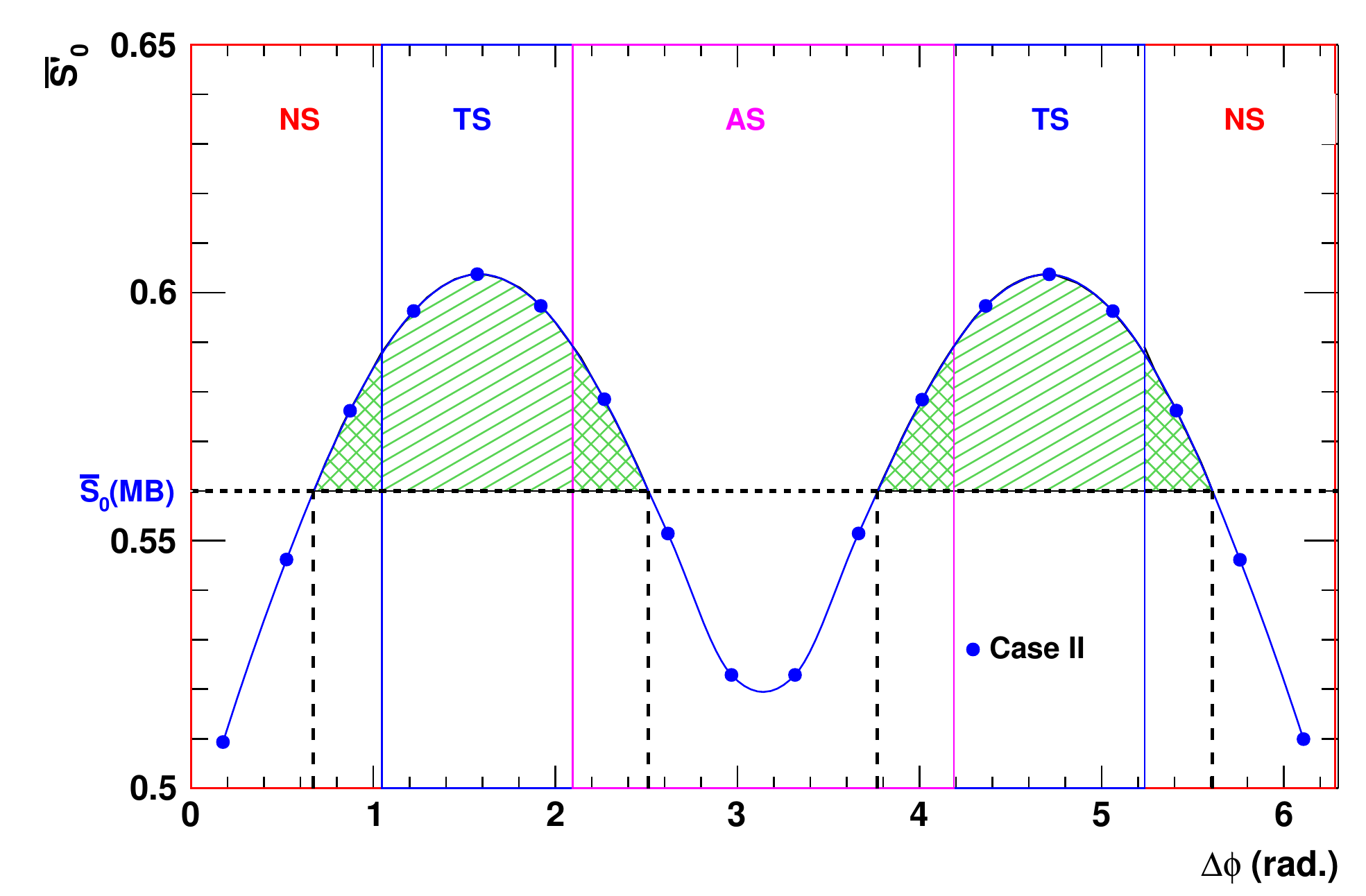}}
	\label{fig:S0SA}
	\caption{Average spherocity ($\bar{S'}_0$)  in $\Delta\phi$ bins are shown for Case I (left) and Case II (right). The horizontal dashed line corresponds to the minimum bias (MB) average spherocity for both cases.}
	\label{fig:S0bar}
\end{figure*}
\subsection{Event shape variations with $\Delta\phi$}
\label{sec:spe}
Geometrical structure of the UE can be investigated by calculating the spherocity distribution as a function of $\Delta\phi$ for both Case I and Case II. We obtained spherocity distributions for each $\Delta\phi$ bin with respect to the leading particle of the event to study the geometrical structure of the angular distributed produced charged particle of the event. Then we plot the average spherocity ($\bar{S'}_{0}$) as a function of $\Delta\phi$ in Fig.~\ref{fig:S0bar}. The average spherocity ($\bar{S}_{0}$) of the MB events is also shown by a black dashed line. In Case I, we observe that the $\bar{S'}_{0}$ increases with $\Delta\phi$ and for the last $\Delta\phi$ bin it gets the same value as MB. Fig.~\ref{fig:S0bar}(a) does not give any fruitful information about the geometrical construction of the UE, apart from that by opening the angle and entering to the Away Side regime, $\bar{S'}_{0}$ presents more isotrope situation relative to the MB. 
The distribution is more clear, in Case II, where we can get an interesting observation with $\Delta\phi$. One can see that $\bar{S'}_{0}$ increases with $\Delta\phi$  and attain higher values (shaded area) than the MB in the TS region which suggest that TS region is more isotropic than the MB (displayed by a horizontal dashed line). One can make an observation here that the charged particle multiplicity in the Transverse Side always lies lower than the MB. We observed that the range of $\Delta\phi$ for TS region is more wider than the traditional definition (see double-shaded area). This study gives the range of the TS from 40$^\circ$ to 140$^\circ$ ($2\pi/9$ to $7\pi/9$) in the one half of the $\Delta\phi$ plane. This suggests that the NS and AS regions are narrower than the traditional definition, thus the UE is in reality  wider than usually considered by about $66\%$ as suggested in Ref.~\cite{Agocs}. This observation is in agreement with~\ref{fig3:Mult}(a), where the normalized multiplicity ($\frac{\rm N_{\rm ch (\Delta\phi)}/\Delta\phi}{\rm N_{\rm ch (MB)}/2\pi}$) crossing 1 (dashed line), suggesting the border between the jet-like, perturbative regime and the bulk, soft UE. This however is around $\Delta\phi \approx \pm $ 0.6 around the Near Side and similarly, the effect is present with less magnitude at the Away Side.

Note, both spherocity and multiplicity show similar structure. One can associate this with the recently-discovered feature of the jet-structure fix point, $R_{fix}$ and the observed jet density scaling, while considering fixed-sized jets in various event multiplicity classes~\cite{Varga:2019,Gemes:2020,Vertesi:2020}.

\begin{figure*}[htbp]
	\centering
	\subfigure[~Case I]{\includegraphics[scale=0.4]{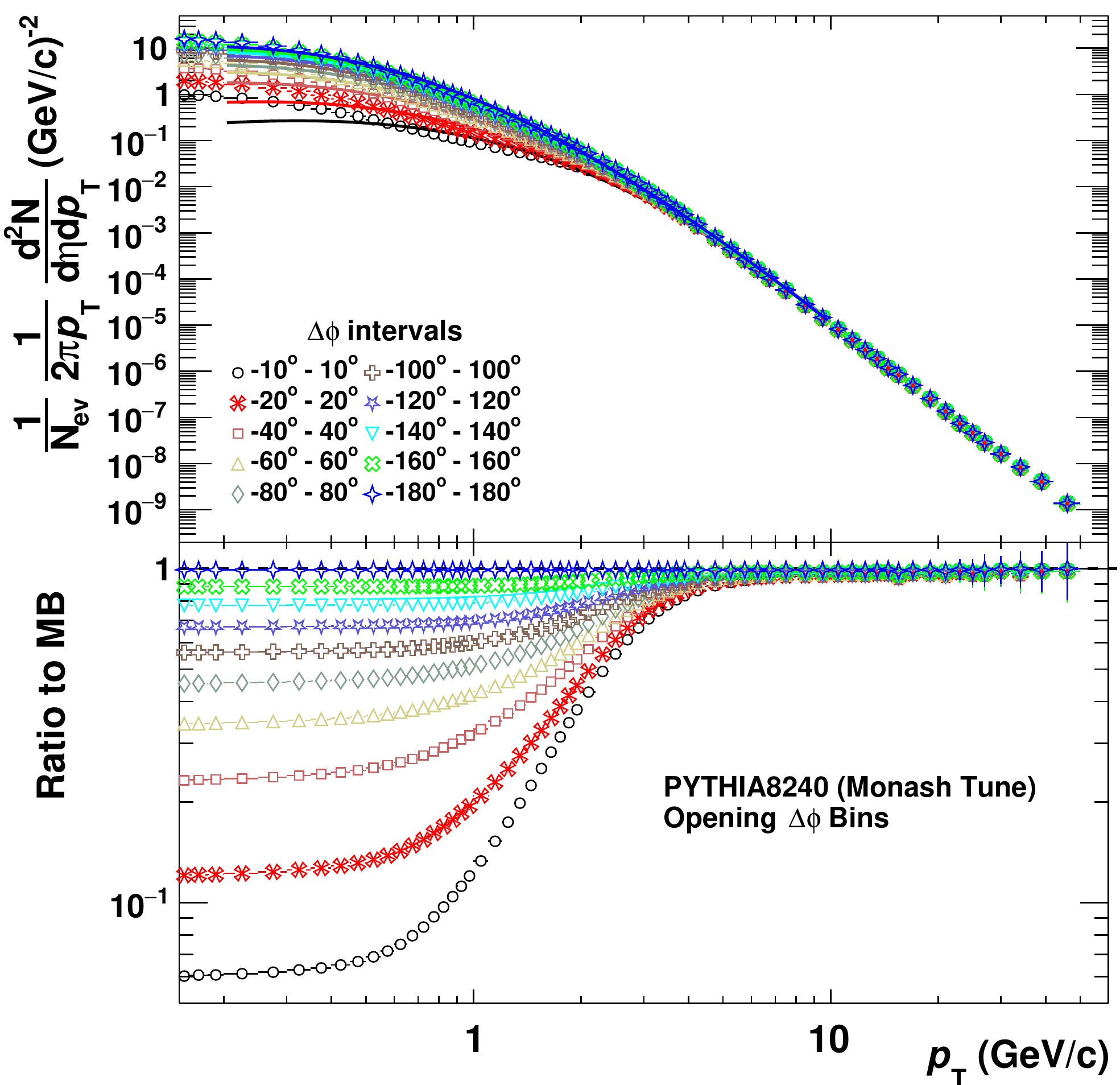}}\quad
	\label{fig2:ptspectraaOA}
	\subfigure[~Case II]{\includegraphics[scale=0.4]{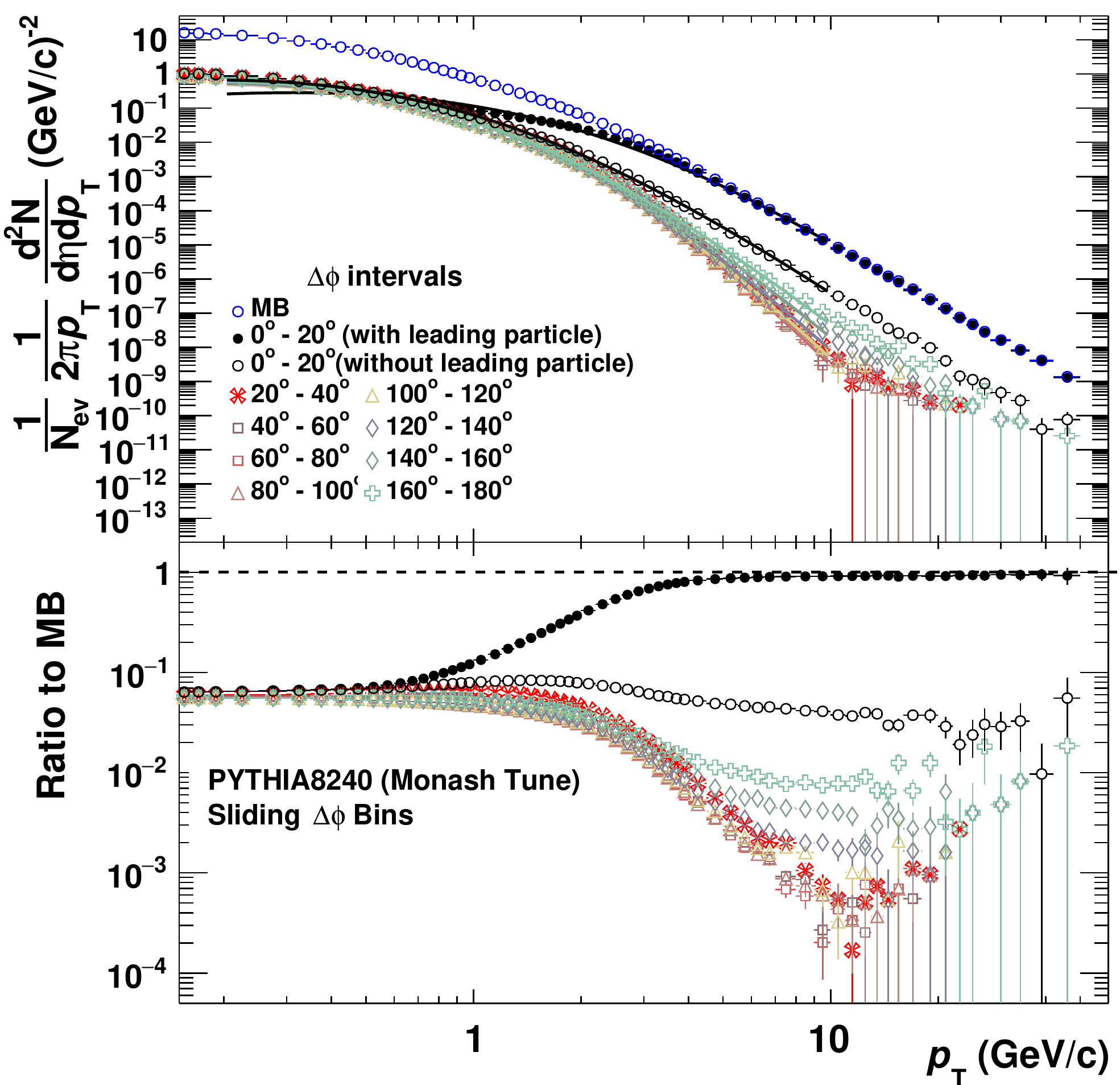}}
	\label{fig2:ptspectraaSA}
	\caption{Charged particle transverse momentum spectra (upper panel) and their ratios to minimum bias spectra (lower panel) are shown for both Case I (left) and Case II (right). First bin of the Case II is plotted for two different scenario: (1) with leading particle (black points) and (2) without leading particle (black circles).}
	\label{fig2:ptspectraa}
\end{figure*}

\subsection{Transverse Momentum Spectra with Non-extensive Tsallis Statistics}
\label{sec:tsallis}
Non-extensive thermodynamics can be applied to high energy physics. In high-energy heavy-ion, and especially in proton-proton collisions we are far from a canonical thermal state, described by the Boltzmann\,--\,Gibbs statistic. More than three decades ago, the Tsallis distribution ~\cite{tsallis} was first proposed as a generalisation of the Boltzmann\,--\,Gibbs distribution which is characterised by an additional parameter $q$. The parameter $q$ measures the deviation from a standard Boltzmann\,--\,Gibbs distribution and known as the measure of `non-extensivity'. Over the past few years, Tsallis distribution has been successful in describing the transverse momentum distributions to high-energy heavy-ion, and especially in proton-proton collisions~\cite{bialas,grigoryan,wong,ristea,biro,parvan1,parvan2,tripathy,zheng1,marques,azmi,sorin,sena,Azmi:2014dwa,Biro:2004qg,Khuntia:2017ite,zheng,de}. 

In the present study, we use one particular form of Tsallis distribution, named Tsallis\,--\,Pareto distributions~\cite{tsbiro:pa2015,tsb:epja2013}, which satisfy the thermodynamic consistency relations and given by: 
\begin{equation}
	f(m_T)=A\cdot \left[1+\frac{q-1}{T_{s}} (m_T-m) \right]^{-\frac{1}{q-1}} \ \ \ ,
	\label{eq:TS}
\end{equation}
where $A$ is scale parameter, $q$ is the non-extensive parameter, $T_{s}$ is a temperature-like parameter, called Tsallis temperature and $m_T=\sqrt{p_T^2+m^2}$ is the transverse mass of the given (identified) hadron species. Note, that we neglect radial flow in this approach, and we use the pion mass for the unidentified charged hadron fits. Parameter, $q$ is manifest as the power of the Tsallis\,--\,Pareto distribution's tail, which can be associated with the power of the transverse momentum spectra in the perturbative (high-$p_T$) regime, $n=\frac{1}{1-q}$.

Experimental observations show that the charged multiplicity in high-energy collisions is dependent on the collision energy~\cite{alice:prl2010} by the following relation:
\begin{equation}
	\frac{\left<{\rm d}N_{ch}/{\rm d}\eta\right>}{\left<N_{part}\right>/2} \propto
	\left\{\begin{array}{ll}
		s_{NN}^{0.15} & \textrm{for AA,} \\
		s_{NN}^{0.11} & \textrm{for pp.} 
	\end{array}\right.
	\label{eq:NmultALICE}
\end{equation}
Using the Eq.~(\ref{eq:NmultALICE}), one can use a given parametric form of identified hadron multiplicity per unit rapidity $\left(\left.\frac{{\rm d}\rm N_{\rm ch }}{{\rm d}y}\right|_{y=0}\right)$ as a function of Tsallis\,--\,Pareto fitting parameters $A$, $T_s$ and $q$ as 
\begin{eqnarray}
	&\left.\frac{\rm dN_{ch}}{\rm dy}\right|_{y=0} = 2\pi A T_s \left[\frac{(2-q)m^2 + 2mT_s + 2T_{s}^2}{(2-q)(3-2q)}\right] \nonumber \\ &  \hspace{2.1truecm} \times\left[1+\frac{q-1}{T_s}m\right]^{-\frac{1}{q-1}}.
	\label{eq:ReCalNmult}
\end{eqnarray}

Since our aim here is to characterise the events by analysing different the UE contributions in the azimuthal space therefore, we fit $p_T$-spectra of charged particle  with Tsallis\,--\,Pareto fitting function and explored the evolution of the Tsallis\,--\,Pareto parameters $A$, $T_{s}$ and $q$ in different $\Delta\phi$ bins for both cases.

The charged particle transverse momentum spectra and their ratio to the MB for proton-proton collisions at $\sqrt{s}$ = 13 TeV using event generator PYTHIA (Monash tune) are shown in  Fig.~\ref{fig2:ptspectraa}. For the Case I, we observe an increase in yield with $\Delta\phi$ at low-$p_T$ ($p_T < 3$ Gev/$c$) while at high-$p_T$ spectra remain nearly unchanged, (see Fig.~\ref{fig2:ptspectraa}(a)) which indicates that most of the high-$p_T$ particle are produced in the first $\Delta\phi$ bin (--10 - 10$^\circ$). ALICE has recently reported that charged particle density shows a sharp rise with the leading charged particle $p_T$ ($p_{T}^{leading}$) at low-$p_T$ ($p_T < 3$ Gev/$c$) and then gets saturated for Transverse Side but Near and Away Side show a slow rise with $p_{T}^{leading}$~\cite{ALICE:vz}. The rise in the Away Side is smaller than the Near Side. As we open more and more the $\Delta\phi$ angle we include Transverse and Away Side regions which have no sufficient high-$p_T$ contributions to change the slope of the spectra at high-$p_T$ ($p_T > 3$ GeV/$c$). This difference at low-$p_T$ can be seen clearly in the lower panel of the Fig.~\ref{fig2:ptspectraa}(a).
In Case II, as we go for higher sliding $\Delta\phi$ bins, we move from Near Side to Transverse Side and finally to Away Side region. Here, we observe that at high-$p_T$ ($p_T > 3$ Gev/$c$) the spectra get softer as we move from Near Side to Transverse Side and again start getting harder as we move to Away Side but never reach to the values observed in the Near Side. The shape of the spectra again confirm that Near Side region contains most of the high-$p_{\rm T}$ particles while Transverse Side which is sensitive to the UE has low-$p_{\rm T}$ particles contributions. At low-$p_T$, the spectra in all three region are almost same which again matches with the ALICE results~\cite{ALICE:vz} where they show that all three region spectra show similar sharp rise with $p_{T}^{leading}$. This difference at high-$p_T$ can be seen clearly in the lower panel of the Fig.~\ref{fig2:ptspectraa}(b). In general, the Case I more sensitive to the low transverse momentum regime, while Case II presents differences for the high and intermediate $\rm p_{\rm T}$ apart from the first bin with leading charged particle.

\subsection{The $\Delta \phi$-scaling of the spectrum-fit parameters}
The $\Delta\phi$ dependent charged particle transverse momentum spectra are then fitted with the  Tsallis\,--\,Pareto function given in Eq.~(\ref{eq:TS}). The fitting parameters $A$, $T_{\rm s}$ and q are shown in Fig.~\ref{fig4:FitParOA} and Fig.~\ref{fig5:FitParSA} for Case I and Case II, respectively. The fitting parameters for both cases are given in Appendix~\ref{apx:TsallisFitPars}. One can make following observations from the fitted Tsallis\,--\,Pareto parameters:
\begin{figure*}[hbt!]
	\centering        
	\vspace*{-0.2cm}
	\includegraphics[scale=0.75]{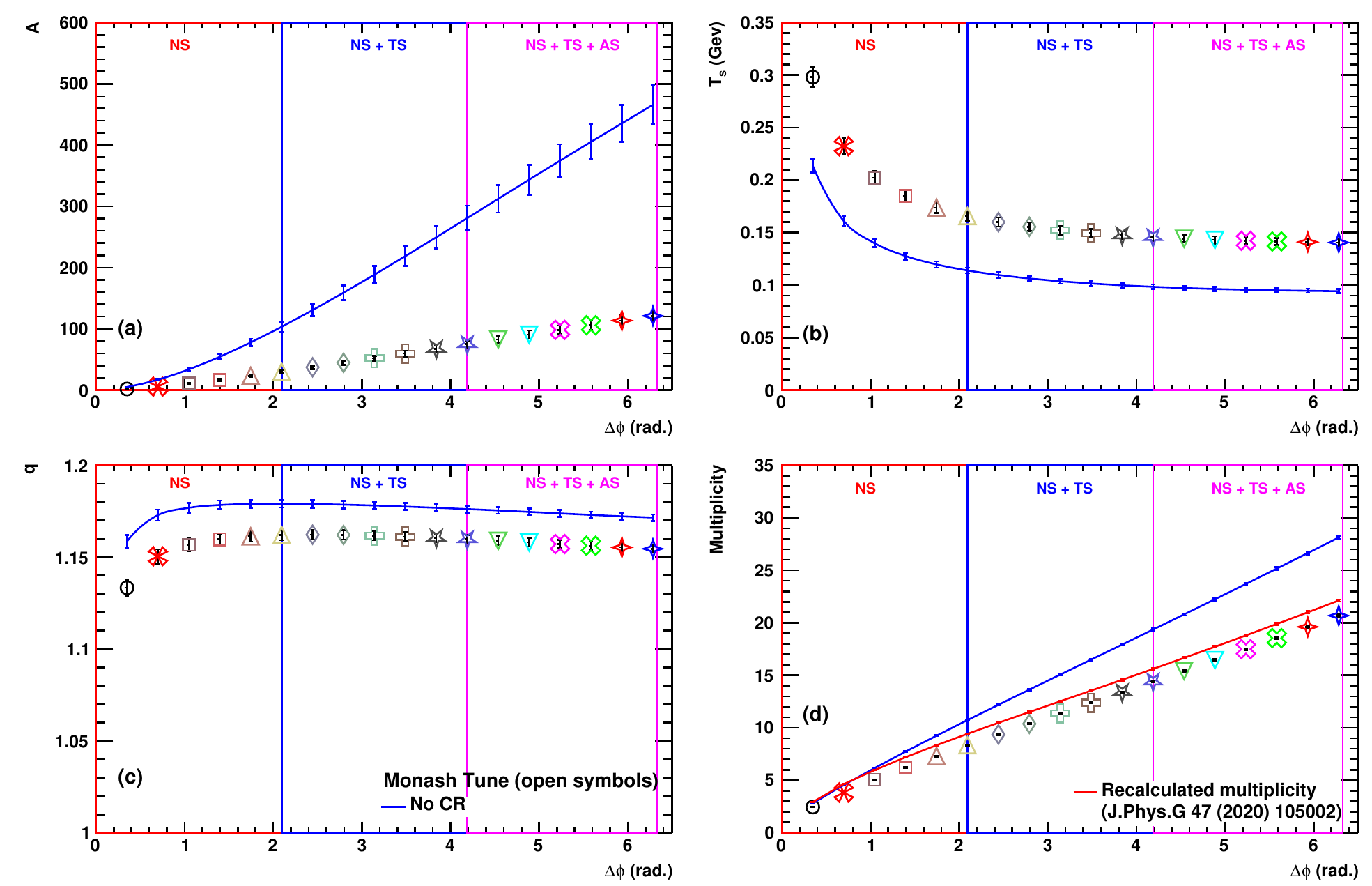}
	\caption{Tsallis\,--\,Pareto scale parameter $A$, Tsallis\,--\,Pareto temperature $T_{s}$ and non-extensive parameter $q$ for pp collisions at $\ensuremath{\sqrt{s}}$ = 13 TeV using PYTHIA8 Monash tune (open markers) and without color reconnection (solid blue line) are shown in Fig.~\ref{fig4:FitParOA}(a), (b) and (c) respectively, for Case I. (d) Multiplicity as a function of $\Delta\phi$ for PYTHIA8 Monash tune (open markers) and without color reconnection (solid blue line) are shown. The recalculated multiplicity values using Eq.~(\ref{eq:ReCalNmult}) are also shown by solid red line.}
	\label{fig4:FitParOA}
\end{figure*}
\begin{figure*}[hbt!]
	\centering        
	\vspace*{-0.2cm}
	\includegraphics[scale=0.75]{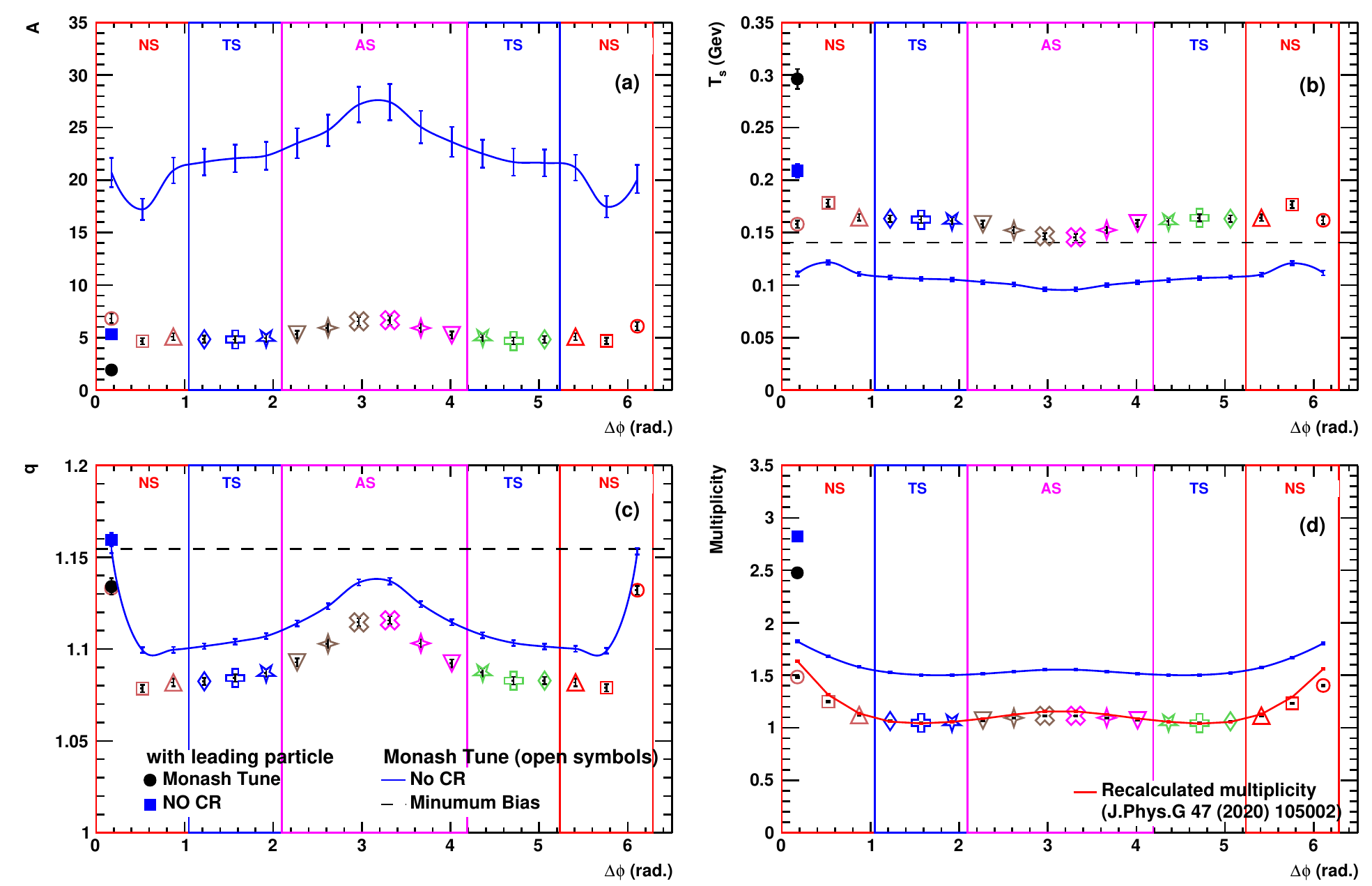}
	\vspace*{-0.5cm}
	\caption{Tsallis\,--\,Pareto scale parameter $A$, Tsallis\,--\,Pareto temperature $T_{s}$ and non-extensive parameter $q$ for pp collisions at $\ensuremath{\sqrt{s}}$ = 13 TeV using PYTHIA8 Monash tune (open markers) and without color reconnection (solid blue line) are shown in Fig.~\ref{fig5:FitParSA}(a), (b) and (c) respectively, for Case II. (d) Multiplicity as a function of $\Delta\phi$ for PYTHIA8 Monash tune (open markers) and without color reconnection (solid blue line) are shown. The recalculated multiplicity values using Eq.~(\ref{eq:ReCalNmult}) are also shown by solid red line.}
	\label{fig5:FitParSA}
\end{figure*}

\begin{description}
	\item[The scale parameter $A$] shows the expected rise with $\Delta\phi$ in Case I because of addition of more charged particles at low-$p_T$ while in Case II it achieves its highest value in Away Side region and have points independent of $\Delta\phi$ in Transverse Side region. We find similar results with the events generated using PYTHIA without Color Reconnection (CR). PYTHIA8 without CR (solid curve) produces more particles than the Monash tune, we get higher values for scale parameter $A$.   
	
	\item[Tsallis\,--\,Pareto temperature, $T_{s}$] in Case I, decreases with $\Delta\phi$ (Fig.~\ref{fig4:FitParOA}(b)). The fall is faster with $\Delta\phi$ for Near Side region but as we add Transverse and Away Side particles this fall gets smaller and become nearly independent of $\Delta\phi$.
	We also observe similar behaviour for Case II shown in Fig.~\ref{fig5:FitParSA}(b). Near Side has the highest $T_{s}$ values and Transverse and Away Side show nearly independent behaviour. Considering the first bin of $\Delta\phi$ without $p_{T}^{leading}$ then $T_{s}$ values lie within 15\% for the full range of $\Delta\phi$. The PYTHIA8 without CR (solid curve) also gives the similar trend with lower values.	
	
	\item[Non-extensivity parameter $q$] is sensitive to the high-$p_T$ part of the spectra. Fig.~\ref{fig4:FitParOA}(c) shows $q$ as a function of $\Delta\phi$ for Case I. Since spectra at high particle momenta, $p_T > 3$ GeV/$c$, are nearly unchanged with $\Delta\phi$ in Case I, $q$ show a small rise in Near Side region and then get saturated with $\Delta\phi$. In Case II, we observe a sharp fall in $q$ from Near Side to Transverse Side then nearly constant in Transverse Side and then a small rise in Away Side with $\Delta\phi$ shown in Fig.~\ref{fig5:FitParSA}(c). The change in $q$ with $\Delta\phi$ is bigger in Case II than the Case I. The PYTHIA8 without CR (solid curve) also result in similar trend but with higher values.
	
	\item[Multiplicity dependence] on the opening/sliding angles is presented on
	Fig.~\ref{fig4:FitParOA}(d)and Fig.~\ref{fig5:FitParSA}(d), respectively. We use Eq.~(\ref{eq:ReCalNmult}) to calculate multiplicity for the different $\Delta\phi$ bins using the $A$, $T_{s}$ and $q$ parameter obtained from the Tsallis\,--\,Pareto fitting~\cite{Biro:2020kve}. This calculation is shown by red color line in Fig.~\ref{fig4:FitParOA}(d) and Fig.~\ref{fig5:FitParSA}(d). Once can see that Eq.~(\ref{eq:ReCalNmult}) works very well for both cases. Further observation is that calculations without
	CR (solid blue lines) generates larger multiplicity, especially getting
	away from the Leading Side.
\end{description} 	

Confirming the validity of Eqs.~(\ref{eq:TS})-(\ref{eq:ReCalNmult}), provides solid basis for the
general idea presented in Ref~\cite{Biro:2020kve}. It seems the model is not only valid in a certain multiplicity bin and at a given c.m. energy,
but fits in all geometrical bins as well. This suggests that Tsallis\,--\,Pareto distribution fits carry information parallel on both the soft
and the hard regimes. Analysing the two cases makes clear, that
Tsallis-parameters represent the properties of the event well, thus
we can apply this in our approach to identify the UE.

\section{Identifying the Underlying Event}
\label{sec:ueid}

Motivated by the observations in the spatial/geometrical structure and momentum space from Figs.~\ref{fig:S0bar}(b) and~\ref{fig2:ptspectraa}, respectively, we try to identify the Underlying Event quantitatively by the Tsallis\,--\,Pareto parameters and by the variation of the investigated global- and event-shape observables.

\begin{figure}[thbp]
	\centering        
	\vspace*{-0.2cm}
	\includegraphics[scale=0.5]{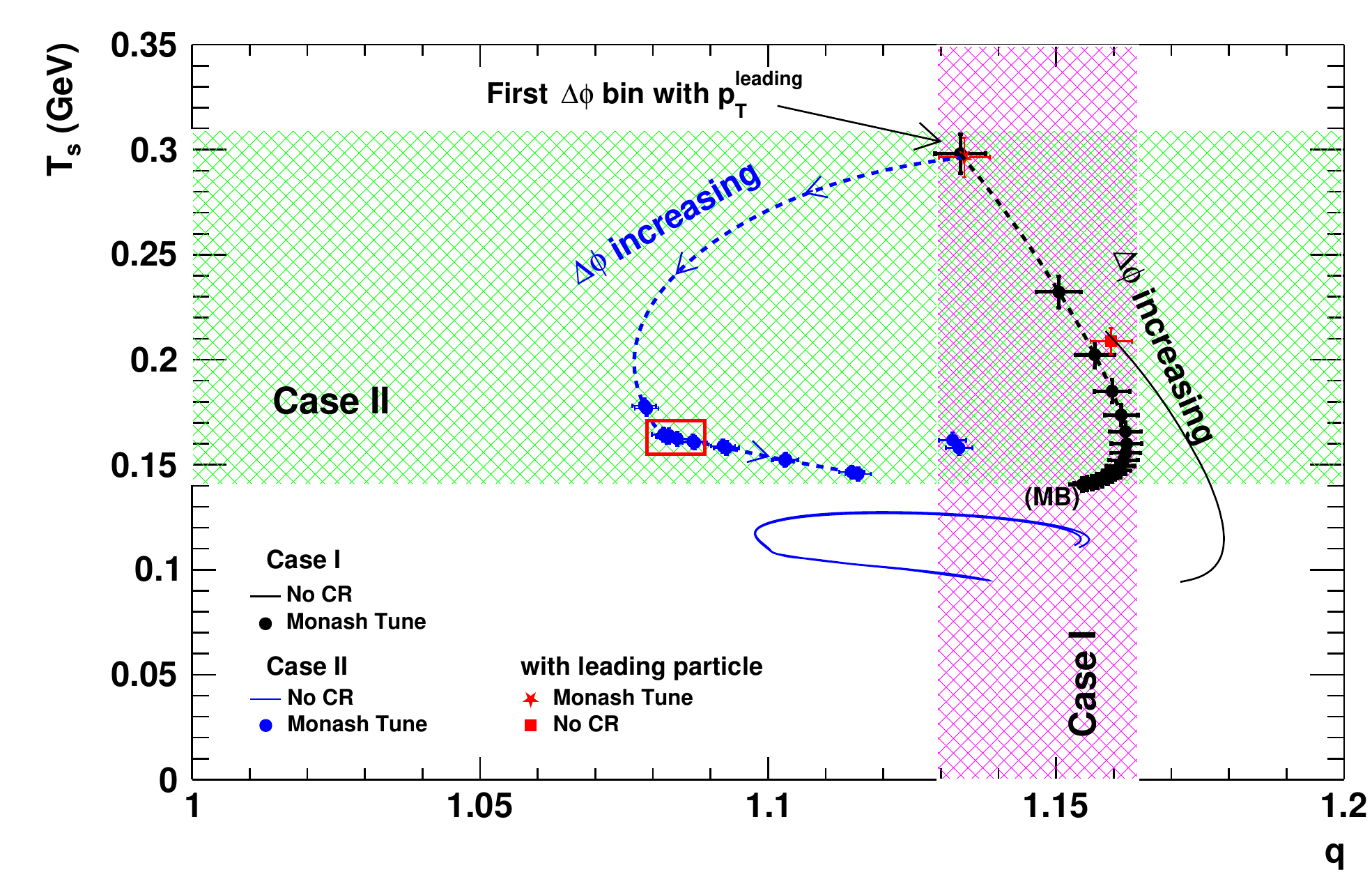}
	\vspace*{-0.5cm}
	\caption{The Tsallis-thermometer, which presents the relation between Tsallis temperature $T_{s}$ and non-extensivity parameter ($q$) for both Case I and Case II. Red box represents the region obtained for new UE definition for Case II as defined in Fig.~\ref{fig:S0bar}(b).}
	\label{fig6:Tvsq}
\end{figure}%

The Tsallis-thermometer, is dedicated to visualize the interconnection between Tsallis temperature ($T_{s}$) and non-extensivity parameter ($q$). This relation presents not only the specific values, but also provides quantitative information on the (non-)extensivity of the system. Since  $T_{s}$ and $q$ are interconnected and depends well on the thermodynamics of the system, investigation of the two parameters is required in parallel. The relation between the $\Delta \phi$-scaling Tsallis temperature and non-extensivity parameter is shown in Fig.~\ref{fig6:Tvsq}. In Case I, with the increase of $\Delta\phi$, $T_{s}$  decreases in contrary to a small change in parameter, $q$. The highest value of $T_s$ corresponds to the first $\Delta\phi$ bin ($-10^{\circ}$ to $10^{\circ}$) while the lowest $T_s$ value were found in the last (full) $\Delta\phi$ bin ($-180^{\circ}$ to $180^{\circ}$). As approaching to the minimum bias, the non-extensitivity inherited from the Near Side is preserved with $q \approx 1.15$, while $T_s$ drops below to its half value due to the increasing multiplicity. This supports the idea, that the evolution in Case I is mainly driven by the increase of the degrees of freedom. (Black points are for Monash tune, black line is with no CR, purple horizontal band is for the parameter region in $q$). 

In Case II, in one hand, once we take into account the leading particle, the first bin is different form the others carrying the highest values for $T_s=0.3$~GeV and $q=1.14$. On the other hand we still observed big variation in $q$ while the change is small in $T_s$ after excluding the first bin ($0^{\circ}$ - 20$^\circ$) with the leading charged particle. (Blue points are for Monash tune, blue line is with no CR, green horizontal band is for the full $T_s$ parameter variation region.) Since in Case II the systems (bins) have the same cone structure, this investigation represent more qualitatively the correlation between the degrees of freedom. Taking the quantitative values of $q$, the lowest values appears in the non-perturbative Transverse Side. Away Side bins has the lowest $T_s$ values, but in the Transverse Side values are a factor of 2 lower, than in the first bin nearby the Leading Particle.

Plotting the corresponding $T_s$ and $q$ values  one can identify a compact locations at $T_s=160\pm 10$~MeV and at $q=1.085 \pm 0.05$ marked as red box in Fig.~\ref{fig6:Tvsq}. Interestingly this region is the lowest in non-extensivity, thus as the most isotrope Transverse Side bins seem to be the closest to the special case of the Tsallis\,--\,Pareto distribution, the $q=1$ Boltzmann\,--\,Gibbs description. Following the obtained geometrical structure from Fig.~\ref{fig:S0bar}, the Underlying Event can be associated with the $[\pm 40^{\circ},\pm 140^{\circ}]$ ($[\pm 2\pi/9,\pm 7\pi/9]$), region in Case II, indeed the more strict CDF-definition in $[\pm 60^{\circ},\pm 120^{\circ}]$ ($[\pm \pi/3,\pm 2\pi/3]$). 

\begin{figure*}[thbp]
	\centering        
	\vspace*{-0.2cm}
	\includegraphics[scale=0.75]{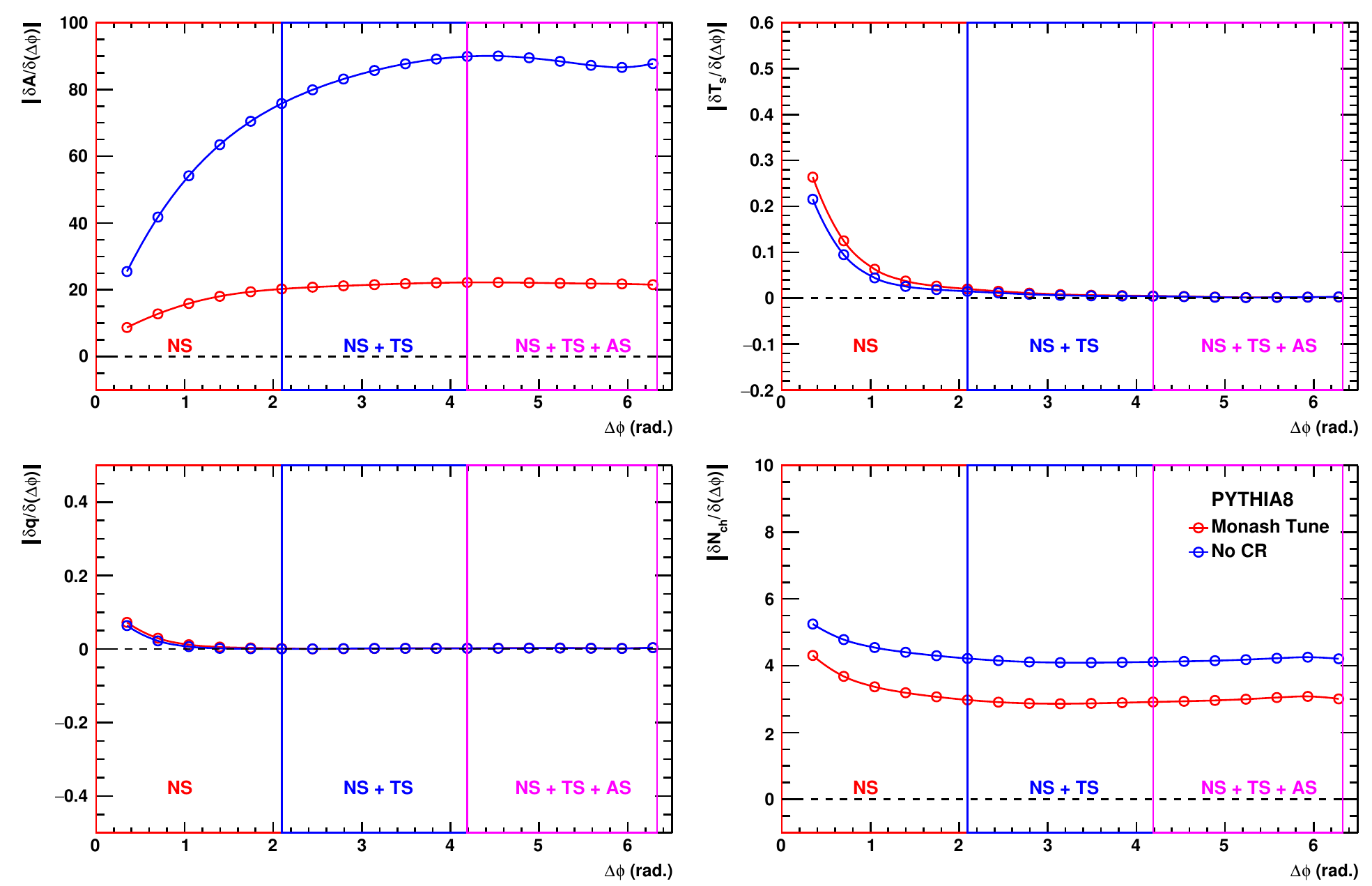}
	\vspace*{-0.5cm}
	\caption{First derivatives (absolute values) of the $\Delta\phi$ dependence of Tsallis\,--\,Pareto scale parameter $A$, Tsallis\,--\,Pareto temperature $T_{s}$, non-extensive parameter $q$ and multiplicity for pp collisions at $\ensuremath{\sqrt{s}}$  = 13 TeV using PYTHIA8 Monash tune (red) and without color reconnection (blue) for Case I.}
	\label{fig:detivativeOA}
\end{figure*}

\begin{figure*}[thbp]
	\centering        
	\vspace*{-0.2cm}
	\includegraphics[scale=0.75]{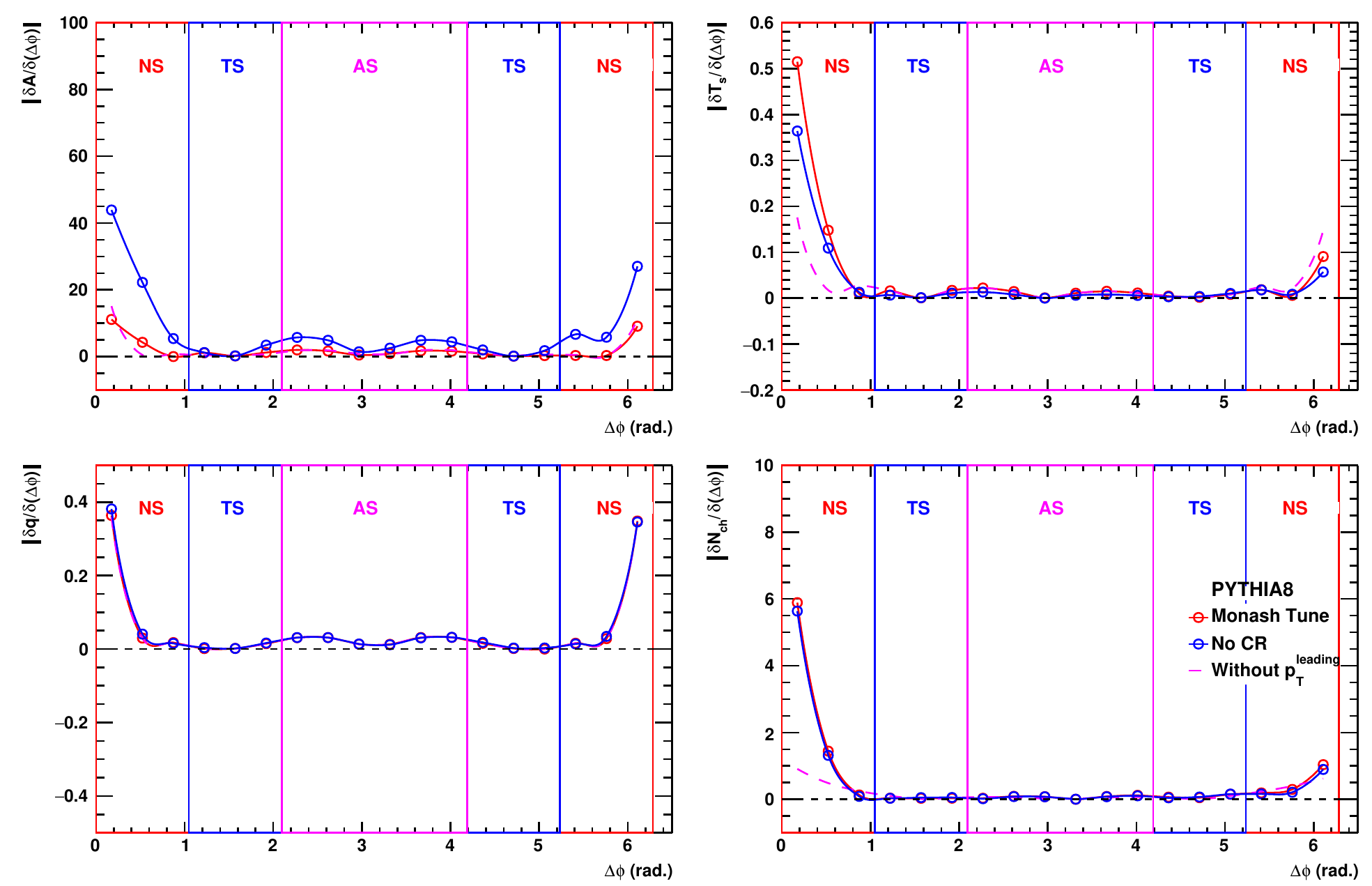}
	\vspace*{-0.5cm}
	\caption{First derivatives (absolute values) of the $\Delta\phi$ dependence of Tsallis\,--\,Pareto scale parameter A, Tsallis\,--\,Pareto temperature $T_{s}$, non-extensive parameter q and multiplicity for pp collisions at $\ensuremath{\sqrt{s}}$ = 13 TeV using PYTHIA8 Monash tune (solid red/dashed magenta lines are with/without leading particle) and without CR (blue) are shown for Case II.}
	\label{fig:detivativeSA}
\end{figure*}

The Tsallis thermometer presents stable  parameter values within the extended UE region. We assume that the variation of the parameters with $\Delta\phi$ carries further information. To understand the geometrical variation of the physical quantities, we investigated the derivatives of the Tsallis\,--\,Pareto scale parameters as a function of $\Delta\phi$ ($\frac{\delta X}{\delta(\Delta\phi)}$,  here  '$X$' can be $A$, $T_{s}$, $q$ and $N_{ch}$) can give us the information about the particle production mechanism in NS, AS and TS. 
We show the absolute value of the first derivatives of the $\Delta\phi$ dependence of Tsallis\,--\,Pareto scale parameter $A$, Tsallis\,--\,Pareto temperature $T_{s}$, non-extensive parameter $q$ and multiplicity for pp collisions at $\ensuremath{\sqrt{s}}$  = 13 TeV using PYTHIA8 Monash tune (red) and PYTHIA8 without color reconnection (blue) for Case I and Case II in  Figs.~\ref{fig:detivativeOA} and~\ref{fig:detivativeSA}, respectively.
For Case I, the first derivative of Tsallis\,--\,Pareto parameters $A$ show change only in first few bins which correspond to the Near Side and then get almost constant with $\Delta\phi$. The $T_s$ and $q$ have nearly zero values of first derivatives except in Near Side. This suggests that adding more and more particles, and increasing the phase-space with opening $\Delta\phi$, from Transverse and Away Side does not change the thermodynamics of the system. As constant multiplicity derivatives present, the system size increase is linear beyond the first bins. On the other hand for Case II, first derivatives of Tsallis\,--\,Pareto parameters are zero in Transverse Side while first derivatives are non-zero for parameters $A$, $T_s$ and $q$ in Near Side and Away Side. The amplitudes of the first derivatives are smaller in Away Side than the Near Side because of the high-$p_T$ contribution in Near Side.  
In Case I, one can conclude from the above discussions that $\Delta\phi$ dependence of the Tsallis\,--\,Pareto parameters $T_s$ and $q$ remain nearly unchanged beyond NS region, which is confirmed by their derivatives having values close to zero, since summing up more-and-more degrees of freedom from the TS and AS washes out the strong effect in the NS, therefore,
\begin{eqnarray}
	\frac{\delta T_{s}}{\delta(\Delta\phi)}     \rightarrow 0 ~~~   \&    ~~~
	\frac{\delta q}{\delta(\Delta\phi)} \rightarrow 0  ~~~~  {\rm ( for ~~\Delta\phi > \pi/3)} \ .
	\label{eq:derivative_TqI}
\end{eqnarray}
Interesting aspect is that, although multiplicity has different slopes for Monash tune and calculations without CR on Fig.~\ref{fig4:FitParOA}, the derivatives, thus the variation of the $T_s$ and $q$ are the same on Fig.~\ref{fig:detivativeOA}.
This support, that the thermodynamics of the system is not changing in Case I apart from the model-dependent "system size", reflected by the variation in the normalization, $A$ and in the multiplicity.

In Case II, we observed changes in the parameter values within the NS and AS region, while in the TS region almost constant figures were found. One can see the zero values for the Tsallis-parameter derivatives in TS regions while non-zero in NS and AS regions as plotted in Fig.~\ref{fig:detivativeSA}, 
\begin{eqnarray}
	\frac{\delta T_{s}}{\delta(\Delta\phi)}    \neq 0  ~~~  \&    ~~
	\frac{\delta q}{\delta(\Delta\phi)}  \neq 0  ~~~~  {\rm  (for~NS~\&~AS) } \\	
	\frac{\delta T_{s}}{\delta(\Delta\phi)}   \approx 0   ~~~  \&    ~~
	\frac{\delta q}{\delta(\Delta\phi)}  \approx 0  ~~~~  {\rm  (for~TS) ~~~~~~~~}
	\label{eq:derivative_TqII}
\end{eqnarray}
In Case II derivatives of the multiplicity and the Tsallis-temperature present zero value outside of the NS on Fig.~\ref{fig:detivativeSA}, which means that the total energy (the product of multiplicity and $T_s$) of a given $\Delta\phi$ bin, should be constant everywhere within the TS and AS regions. If both total energy and phase-space are equally distributed outside the NS, then transverse momentum spectra is the one, which carries information on the thermodynamics. Furthermore, the tail of the $p_T$ distribution is connected to the $q$ value, which relation can provide a bin-by-bin measure to describe quantitatively the hadron production of the Underlying Event.  

The conclusion from the above discussions is that charged particle production in the UE region is fully interconnected with the Transverse Side, and since it does not change much in the neighbouring $\Delta\phi$ bins, the extension of the UE presented in Fig.~\ref{fig:S0bar} is advocated. In these widened regions due to the stability of the multiplicity, the Tsallis parameters reflect the UE's limits well. The constancy of the  Tsallis parameters $A$, $T_s$, $q$, and multiplicity in the TS, can be interpreted in the two cases: (i) the linear increase of the system size in Case I by washing out the strong effect at the NS, and (ii) selecting constant total energy bins as it is given in Case II.  

\section{Corss-check with Spherocity}

\begin{figure*}[thbp]
	\centering        
	\vspace*{-0.2cm}
	\includegraphics[scale=0.75]{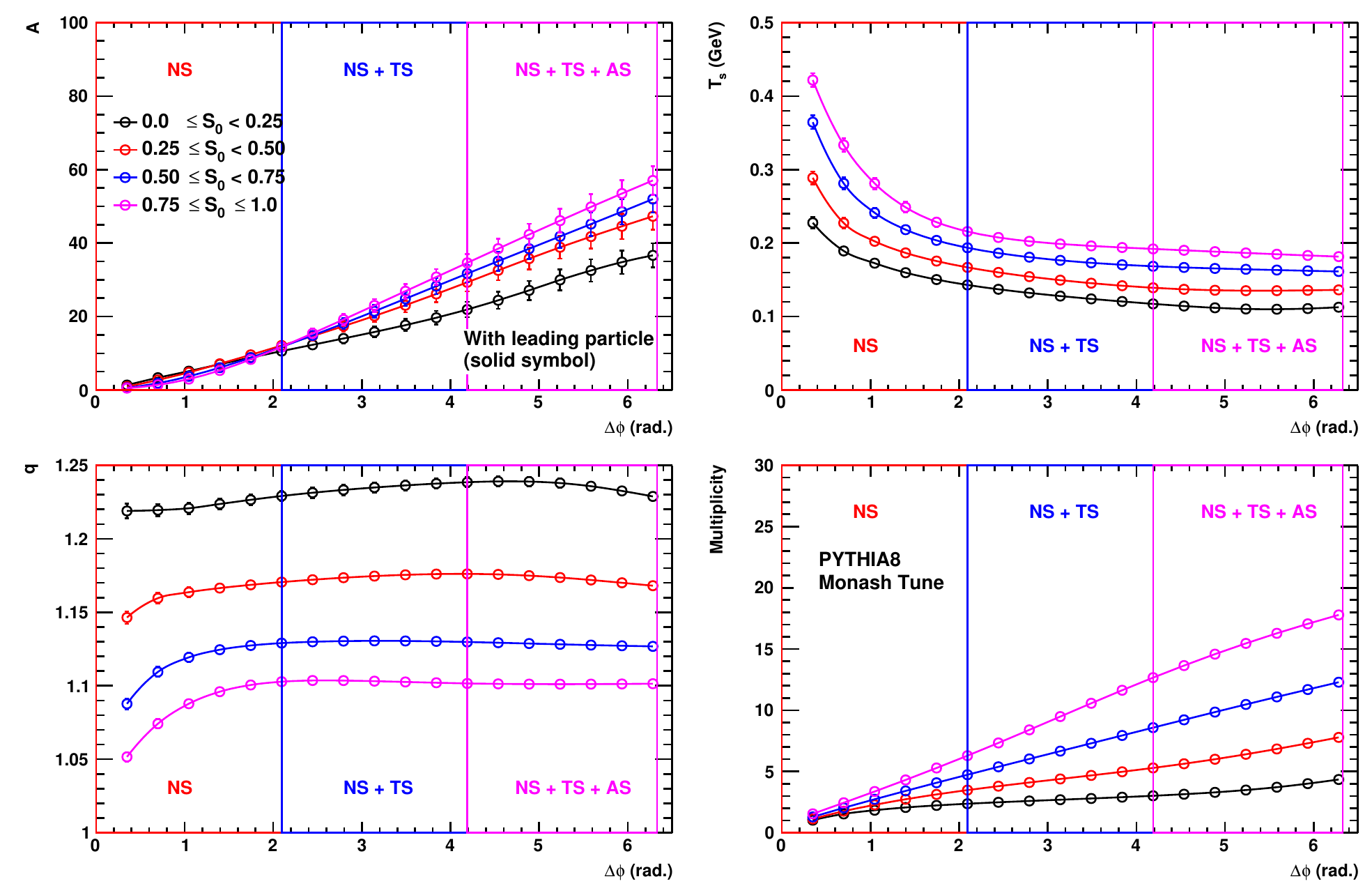}
	\vspace*{-0.5cm}
	\caption{$\Delta\phi$ dependence of Tsallis\,--\,Pareto parameters ($A$, $T_{S}$ and $q$) and multiplicity for different spherocity classes in pp collisions at $\ensuremath{\sqrt{s}}$  = 13 TeV using PYTHIA Monash tune are shown for Case I.}
	\label{fig:ParOAS0}
\end{figure*}

\begin{figure*}[thbp]
	\centering        
	\vspace*{-0.2cm}
	\includegraphics[scale=0.75]{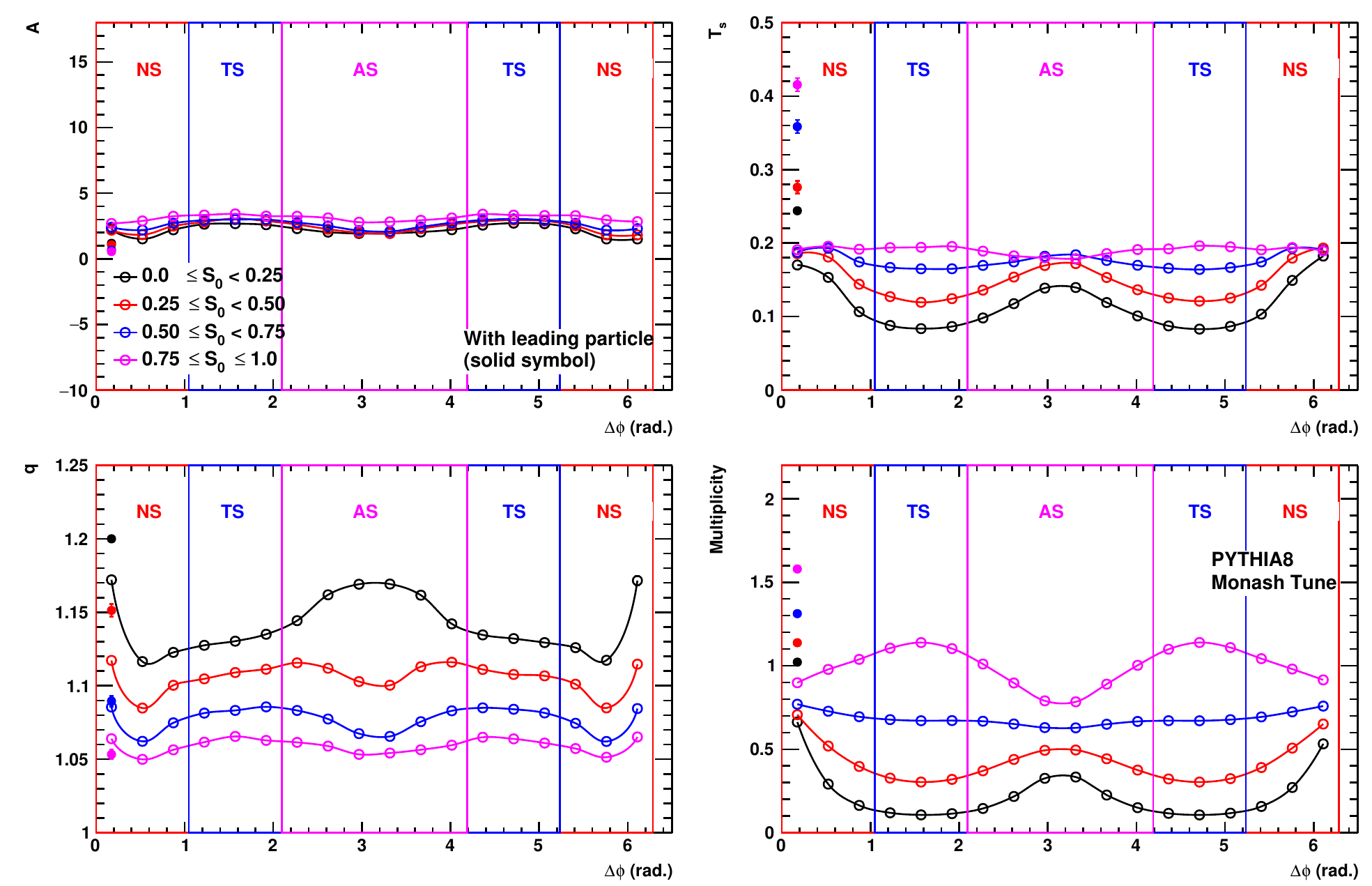}
	\vspace*{-0.5cm}
	\caption{$\Delta\phi$ dependence of Tsallis\,--\,Pareto parameters ($A$, $T_{S}$ and $q$) and multiplicity for different spherocity classes in pp collisions at $\ensuremath{\sqrt{s}}$ = 13 TeV using PYTHIA Monash tune are shown for Case II.}
	\label{fig:ParSAS0}
\end{figure*}

Spherocity values carry compound information on the angular structure and the (absolute value) distribution of the transverse momenta vector. It is easy to imagine, that the $p_T$ distribution and the Tsallis-parameters of the $\Delta\phi$ correlations will be different for "jetty" or "isotrope" event selections. Therefore the obtained Tsallis-parameter values will vary in different spherocity classes. A possible cross-check of our presented results is to compare the quantified underlying-event properties by applying event-shape-variable classified events. To investigate this we divide $S_{0}$ in four regions (0$-$0.25, 0.25$-$0.50, 0.50$-$0.75 and 0.75$-$1.0) to select different kind of events for further study. We fit Tsallis\,--\,Pareto function to the particle $p_T$ spectra for all four $S_{0}$ regions for Case I and Case II. The obtained Tsallis parameters for different $\Delta\phi$ are plotted for Case I and Case II in Fig.~\ref{fig:ParOAS0} and Fig.~\ref{fig:ParSAS0}, respectively. Note, the averages of these detailed results are equivalent to the proper minimum bias curves in (Figs.~\ref{fig4:FitParOA} and~\ref{fig5:FitParSA}), where no $S_0$ selection were applied.
In Case I, $T_{s}$ and $q$ show a similar trend with $\Delta\phi$ as observed in the minimum bias. Following the expectations $T_{s}$ gets the highest values for jetty events and the lowest for isotropic events while $q$ gets its highest values for isotropic events and lowest for jetty events ($\delta q = 0.15$). The multiplicity always increases with $\Delta\phi$ and its slope is higher for isotropic events than the jetty one. One can see that multiplicity has the same values in the first few $\Delta\phi$ bins for all four sections of the $S_{0}$ and then multiplicity values rises faster in isotropic events than the jetty ones. 
Since jetty events have more contribution from the NS, and less from the TS, the multiplicity values are flat for this class. This behaviour of multiplicity is shown in Fig.~\ref{fig:ParOAS0}(d). The derivatives of the Tsallis parameters with $\Delta\phi$ are the largest for the isotrope event classes and  parameters present minimal variation for jetty events.

In Case II one can observe that isotrope events achieve the highest $T_{s}$ in the first bin of $\Delta\phi$ and then start decreasing when it enters in TS and then rise again a bit in AS. Interestingly the change in $T_{s}$ become almost zero for jetty events, since there is no relevant contribution from the UE relative to the NS. This is in agreement with our earlier observation presented in the Tsallis-thermometer, Fig.~\ref{fig6:Tvsq}. The $\frac{\delta T_{s}}{\delta(\Delta\phi)} \rightarrow 0$ confirms this behaviour for jetty events more 

The change in parameter $q$ shows also $\delta q \approx 0.15$ decrease while going from jetty to isotropic events. 
Further observation on multiplicity: as we move from jetty to isotropic events, multiplicity is rising for isotrope event ($0.75 < S_{0} < 1.0$) and decreasing for jetty ones ($0.0 < S_{0} < 0.25$). Is is interesting to see, that the variation of this effect is the strongest in the TS.

In summary we can confirm, that quantified Underlying Event properties by the Tsallis\,--\,Pareto fits from jetty to isotrope events, correlates with our expectation of the geometrical distribution within an event as we have obtained in Section~\ref{sec:ueid}.

\section{Summary}
We investigated the geometrical structure of charged particle production in proton-proton collisions at $\sqrt{s}$ = 13 TeV using PYTHIA8 Monte Carlo event generator. The properties of the Underlying Event have been quantified in angular bins, which led us to explore the event structure. In parallel the multiplicity and midrapidity transverse momentum spectra of charged hadrons have been analyzed. We used the non-extensive statistical approach for our study.  We quantified the properties of the event slices in two different angular selection cases to identify the Underlying Event. Case I provided a general trend summing up the hadronic degrees of freedoms as opening the phase-space from the leading particle, while Case II described the geometrical structure within equal, $20^{\circ}$ angular bin slices.

We found that the standard definition of the Underlying Event can be extended with about 66\% in geometry within the range, $[\pm 40^{\circ},\pm 140^{\circ}]$ in comparison to the widely-used CDF-definition, $[\pm 60^{\circ},\pm 120^{\circ}]$. Indeed we found an upper transverse-momentum threshold, $p_T \lesssim$ 3-4~GeV/$c$ for hadrons with Underlying Event origin. Location of the Underlying Event on the Tsallis-thermometer has been also presented at $T_s=160\pm 10$~MeV and at $q=1.085 \pm 0.05$. Non-extensivity value was found to be the closest to the Boltzmann\,--\,Gibbs limit ($q=1.0$) of the Tsallis\,--\,Pareto distribution's in isotropic events.

These findings were cross-checked with the parameter derivatives and spherocity-classified events. The obtained, nearly-zero angular variation of the spectral parameters in the Transverse Side region can support the identification and better localization of the Underlying Event by the Tsallis\,--\,Pareto parameters. Finally our quantified values correlated well with the spherocity-classified parameter trends.

\section*{Acknowledgements}
Authors would like to thank the Hungarian National Research, Development and Innovation Office (NKFIH) under the contract numbers OTKA K135515, K123815 and NKFIH 2019-2.1.11-TET-2019-00078, 2019-2.1.11-TET-2019-00050, Wigner GPU Laboratory, Mexican Conacyt project CF2042 and the support of the Coordinacion de la Investigacion Cientifica, Mexico. Discussion on the fit procedure with G.~B\'\i r\'o and T.S.~Bir\'o is highly appreciated.

\newpage
\appendix

\section{Tsallis\,--\,Pareto fitting parameters}
\label{apx:TsallisFitPars}

\begin{table*}[h!]
	\caption{Tsallis\,--\,Pareto fitting parameters for the fitting of PYTHIA (Monash tune) spectra at $\sqrt{s}=13$ TeV in pp collisions for Case I}
	\centering
	\begin{tabular}{c c c c c}
		\hline
		{\bf $\Delta\phi$ bin} & {\bf A} & {\bf ${\rm \bf T}_{s}$ (GeV)} & {\bf q} & {\bf $\bf \chi^{2}/ndf$} \\ \hline \hline
		
		-10$^\circ$ --  10$^\circ$ &   1.759${\displaystyle \pm }$0.133 &  0.298${\displaystyle \pm }$0.009 & 1.133${\displaystyle \pm }$0.004 & 2.511 \\
		-20$^\circ$ --  20$^\circ$ &   5.540${\displaystyle \pm }$0.479 &  0.232${\displaystyle \pm }$0.007 & 1.150${\displaystyle \pm }$0.004 & 1.975 \\		
		-30$^\circ$ --  30$^\circ$ &   10.573${\displaystyle \pm }$0.903 & 0.202${\displaystyle \pm }$0.006 & 1.157${\displaystyle \pm }$0.003 & 1.808 \\		
		-40$^\circ$ --  40$^\circ$ &   16.547${\displaystyle \pm }$1.386 & 0.185${\displaystyle \pm }$0.005 & 1.159${\displaystyle \pm }$0.003 & 1.646 \\		
		-50$^\circ$ --  50$^\circ$ &   23.091${\displaystyle \pm }$1.904 & 0.174${\displaystyle \pm }$0.005 & 1.161${\displaystyle \pm }$0.003 & 1.582 \\		
		-60$^\circ$ --  60$^\circ$ &   30.027${\displaystyle \pm }$2.443 & 0.166${\displaystyle \pm }$0.005 & 1.162${\displaystyle \pm }$0.003 & 1.431 \\ \hline		
		-70$^\circ$ --  70$^\circ$ &   37.195${\displaystyle \pm }$2.991 & 0.159${\displaystyle \pm }$0.004 & 1.162${\displaystyle \pm }$0.003 & 1.449  \\	
		-80$^\circ$ --  80$^\circ$ &   44.544${\displaystyle \pm }$3.544 & 0.156${\displaystyle \pm }$0.004 & 1.162${\displaystyle \pm }$0.003 & 1.209 \\		
		-90$^\circ$ --  90$^\circ$ &   52.005${\displaystyle \pm }$4.093 & 0.152${\displaystyle \pm }$0.004 & 1.162${\displaystyle \pm }$0.002 & 1.099 \\		
		-100$^\circ$ --  100$^\circ$ & 59.586${\displaystyle \pm }$4.642 & 0.149${\displaystyle \pm }$0.004 & 1.161${\displaystyle \pm }$0.002 & 1.081 \\		
		-110$^\circ$ --  110$^\circ$ & 67.257${\displaystyle \pm }$5.186 & 0.147${\displaystyle \pm }$0.004 & 1.160${\displaystyle \pm }$0.002 & 1.065 \\		
		-120$^\circ$ --  120$^\circ$ & 74.980${\displaystyle \pm }$5.725 & 0.146${\displaystyle \pm }$0.004 & 1.159${\displaystyle \pm }$0.002 & 1.097 \\	\hline		
		-130$^\circ$ --  130$^\circ$ & 82.733${\displaystyle \pm }$6.256 & 0.144${\displaystyle \pm }$0.003 & 1.159${\displaystyle \pm }$0.002 & 1.053 \\ 	
		-140$^\circ$ --  140$^\circ$ & 90.482${\displaystyle \pm }$6.778 & 0.143${\displaystyle \pm }$0.003 & 1.158${\displaystyle \pm }$0.002 & 1.047 \\		
		-150$^\circ$ --  150$^\circ$ & 98.206${\displaystyle \pm }$7.292 & 0.142${\displaystyle \pm }$0.003 & 1.157${\displaystyle \pm }$0.002 & 1.045 \\		
		-160$^\circ$ --  160$^\circ$ & 105.856${\displaystyle \pm }$7.795 & 0.142${\displaystyle \pm }$0.003 & 1.156${\displaystyle \pm }$0.002 & 1.030 \\		
		-170$^\circ$ --  170$^\circ$ & 113.428${\displaystyle \pm }$8.289 & 0.141${\displaystyle \pm }$0.003 & 1.155${\displaystyle \pm }$0.002 & 0.987 \\	
		-180$^\circ$ --  180$^\circ$ & 121.037${\displaystyle \pm }$8.784 & 0.141${\displaystyle \pm }$0.003 & 1.155${\displaystyle \pm }$0.002 & 0.924 \\	
		
	\end{tabular}
	\label{table1:OA}
\end{table*}	
\begin{table*}[h!]
	\caption{Tsallis\,--\,Pareto fitting parameters for the fitting of PYTHIA (Monash tune) spectra at $\sqrt{s}=13$ TeV in pp collisions for Case II }
	\centering
	\begin{tabular}{ c c c c c }
		\hline
		{\bf $\Delta\phi$ bin} & {\bf A} & {\bf ${\rm \bf T}_{s}$ (GeV)} & {\bf q} & {\bf $\bf \chi^{2}/ndf$} \\ \hline \hline
		0$^\circ$ --  20$^\circ$  (with $p_{\mathrm{T}}^{\mathrm{leading}}$)  &  1.912${\displaystyle \pm }$0.152 & 0.296${\displaystyle \pm }$0.009 & 1.134${\displaystyle \pm }$0.004 &	2.471	\\
		0$^\circ$ --  20$^\circ$  (without $p_{\mathrm{T}}^{\mathrm{leading}}$)  &  6.814${\displaystyle \pm }$0.456 & 0.158${\displaystyle \pm }$0.003 & 1.133${\displaystyle \pm }$0.002 & 0.968	\\
		20$^\circ$ --  40$^\circ$   &  4.663${\displaystyle \pm }$0.300 & 0.178${\displaystyle \pm }$0.003 & 1.078${\displaystyle \pm }$0.002 & 2.044	\\
		40$^\circ$ --  60$^\circ$   &  5.105${\displaystyle \pm }$0.336 & 0.164${\displaystyle \pm }$0.003 & 1.082${\displaystyle \pm }$0.002 & 1.792	\\  \hline
		60$^\circ$ --  80$^\circ$   &  4.852${\displaystyle \pm }$0.329 & 0.163${\displaystyle \pm }$0.003 & 1.082${\displaystyle \pm }$0.002 & 2.111	\\ 
		80$^\circ$ --  100$^\circ$   & 4.825${\displaystyle \pm }$0.334 & 0.162${\displaystyle \pm }$0.003 & 1.084${\displaystyle \pm }$0.002 & 2.152	\\
		100$^\circ$ --  120$^\circ$   & 4.959${\displaystyle \pm }$0.345 & 0.161${\displaystyle \pm }$0.003 & 1.087${\displaystyle \pm }$0.002 & 2.230	\\ \hline
		120$^\circ$ --  140$^\circ$   & 5.313${\displaystyle \pm }$0.365 & 0.157${\displaystyle \pm }$0.003 & 1.093${\displaystyle \pm }$0.002 & 1.928	\\
		140$^\circ$ --  160$^\circ$   & 5.947${\displaystyle \pm }$0.405 & 0.152${\displaystyle \pm }$0.003 & 1.103${\displaystyle \pm }$0.002 & 1.779	\\ 
		160$^\circ$ --  180$^\circ$   & 6.565${\displaystyle \pm }$0.440 & 0.146${\displaystyle \pm }$0.003 & 1.114${\displaystyle \pm }$0.002 & 1.254	\\
		180$^\circ$ --  200$^\circ$   & 6.678${\displaystyle \pm }$0.449 & 0.145${\displaystyle \pm }$0.003 & 1.116${\displaystyle \pm }$0.002 & 1.355	\\
		200$^\circ$ --  220$^\circ$   & 5.935${\displaystyle \pm }$0.403 & 0.152${\displaystyle \pm }$0.003 & 1.103${\displaystyle \pm }$0.002 & 1.726	\\ 
		220$^\circ$ --  240$^\circ$   & 5.249${\displaystyle \pm }$0.357 & 0.159${\displaystyle \pm }$0.003 & 1.092${\displaystyle \pm }$0.002 & 1.852	\\ \hline
		240$^\circ$ --  260$^\circ$   & 5.014${\displaystyle \pm }$0.348 & 0.160${\displaystyle \pm }$0.003 & 1.087${\displaystyle \pm }$0.002 & 2.145	\\
		260$^\circ$ --  280$^\circ$   & 4.693${\displaystyle \pm }$0.329 & 0.164${\displaystyle \pm }$0.003 & 1.083${\displaystyle \pm }$0.002 & 2.375	\\
		280$^\circ$ --  300$^\circ$   & 4.841${\displaystyle \pm }$0.333 & 0.163${\displaystyle \pm }$0.003 & 1.083${\displaystyle \pm }$0.002 & 2.182	\\ \hline
		300$^\circ$ --  320$^\circ$   & 5.106${\displaystyle \pm }$0.340 & 0.164${\displaystyle \pm }$0.003 & 1.082${\displaystyle \pm }$0.002 & 1.891	\\ 
		320$^\circ$ --  340$^\circ$   & 4.698${\displaystyle \pm }$0.302 & 0.177${\displaystyle \pm }$0.003 & 1.079${\displaystyle \pm }$0.002 & 1.960	\\
		340$^\circ$ --  360$^\circ$   & 6.093${\displaystyle \pm }$0.405 & 0.162${\displaystyle \pm }$0.003 & 1.131${\displaystyle \pm }$0.002 & 1.041	\\  \hline	
		
	\end{tabular}
	\label{table2:SA}
\end{table*}

\end{document}